\documentclass[%
twocolumn,amsmath,amssymb
reprint,
superscriptaddress,
aps,
]{revtex4-1}
\usepackage{xcolor}
\usepackage{graphicx}
\usepackage{dcolumn}
\usepackage{bm}
\usepackage{float}
\usepackage{threeparttable}
\usepackage{upquote}  
\usepackage{multirow}
\usepackage[colorlinks=true,linkcolor=red]{hyperref}%

\usepackage[mathlines]{lineno}

\usepackage{makecell}

\def\beq{\begin{equation}}
\def\eeq{\end{equation}}
\def\bea{\begin{eqnarray}}
\def\eea{\end{eqnarray}}
\def\brcl{\begin{array}{rcl}}
\def\bccl{\begin{array}{ccl}}
\def\blcl{\begin{array}{lcl}}
\def\err{\end{array}}

\def\fatx{{\bf x}}

\renewcommand{\textcolor}[2]{#2}

\begin{document}

\preprint{APS/123-QED}

\title{{\em Ab initio} machine learning in chemical compound space}

\author{Bing Huang}
\author{O. Anatole von Lilienfeld}
\email{anatole.vonlilienfeld@univie.ac.at}
\affiliation{University of Vienna, Faculty of Physics, 1090 Vienna, Austria}
\affiliation{Institute of Physical Chemistry and National Center for Computational Design and Discovery of Novel Materials (MARVEL), Department of Chemistry, University of Basel, 4056 Basel, Switzerland}

\date{\today}

\begin{abstract}
Chemical compound space (CCS), the set of all theoretically conceivable combinations of chemical elements and (meta-)stable geometries that make up matter, is colossal. The first principles based virtual sampling of this space, for example in search of novel molecules or materials which exhibit desirable properties, is therefore prohibitive for all but the smallest sub-sets and simplest properties.
We review studies aimed at tackling this challenge using modern machine learning techniques based on (i) synthetic data, typically generated using quantum mechanics based methods, and (ii) model architectures inspired by quantum mechanics. 
Such Quantum mechanics based Machine Learning (QML) approaches combine the numerical efficiency of statistical surrogate models with an {\em ab initio} view on matter. 
\textcolor{red}{They} rigorously reflect the underlying physics in order to reach universality and transferability across CCS. 
While state-of-the-art approximations to quantum problems impose severe computational bottlenecks, recent QML based developments indicate the possibility of substantial acceleration without sacrificing the predictive power of quantum mechanics. 
\end{abstract}

\pacs{Valid PACS appear here}
\maketitle

\tableofcontents

\section{Introduction}
Promising applications of machine learning techniques have been rapidly gaining momentum throughout the chemical sciences.
Apart from this present special issue in {\em Chemical Reviews}, 
a number of special issues in common theoretical chemistry community journals
have appeared, including {\em Int. J. Quantum Chem.} (2015)~\cite{RuppForeword2015},
{\em J. Chem. Phys.} (2018)~\cite{rupp2018guest}, 
{\em J. Phys. Chem.} (2018)~\cite{Guo2020editorial}, 
{\em J. Phys. Chem. Lett.} (2020)~\cite{prezhdo2020editorial},
or {\em Nature Communications} (2020)~\cite{tkatchenko2020editorial}.
Books, essays, reviews, and opinion pieces have also
been contributed by practitioners in the field~\cite{MachineLearningMeetsQuantumPhysics2020book,RaghusReview2016,QMLessayAnatole,kitchin2018MLinCatalysis,huang2018quantum,Walsh2018MLnature,aspurulindhreiher2018revolution,faber2019modeling,batista2019designMLalchemy,Anatole2020NatureReview,Burke2020retrospective,noe2020review,faber2020lecturenotesphysics,MachineLearningMeetsQuantumPhysics2020book,unke2020MLforFFs,AspuruOpreaRoitbergTetko2020qsar,chibani2020machine,dral2020quantum,haghighatlari_learning_2020}. 
Such growth of interest prompted a general discussion in {\em Angewandte Chemie} within a trilogy of essays by Hoffmann and Malrieu on the seemingly conflicting nature of simulation and understanding in quantum chemistry~\cite{RoaldHoffmann2020essayA,RoaldHoffmann2020essayB,RoaldHoffmann2020essayC}.
The overall enthusiasm in the hard sciences for machine learning has even led to the introduction of novel journals, such as Springer's {\em Nature Machine Intelligence}, 
IOP's {\em Machine Learning: Science and Technology}~\cite{von2020introducing},
or Wiley's
{\em Applied Artificial Intelligence Letters}~\cite{pyzer2020editorialAIIL}.

In this review, we attempt to provide a comprehensive overview on recent progress made regarding the problem of using machine 
learning models to train and predict quantum properties throughout chemical compound space (CCS). 
In contrast to the current trend of machine learning in quantum computing, 
we here refer to the application of statistical learning to quantum properties as `quantum machine learning' (QML). 
This notation follows a common convention in atomistic simulation where the quantum nature of the object to be studied corresponds to 
a prefix, while the actual algorithms are rather classical in nature.
Examples include Quantum Monte Carlo or Quantum Molecular Dynamics (also known as {\em ab initio} or `first principles' molecular dynamics).

Within this introductory section we will begin by first providing a qualitative description of chemical compound space (CCS) in terms of fundamental variables which is consistent with the quantum mechanical picture within the Born-Oppenheimer approximation and neglecting nuclear quantum and relativistic effects. 
Thereafter, we briefly review related but complementary and system specific QML models which predominantly are not used throughout CCS but rather for training and predicting potential energies and forces in terms of conformational degrees of freedom, e.g.~using molecular dynamics.  
Quantum mechanics based explorations for the purpose of materials design are mentioned subsequently, followed by a short subsection on studies which establish the quantitative and rigorous quantum chemistry based view on CCS. 

\subsection{Multi-scale nature of CCS}
Figuratively speaking, CCS refers to the virtual set of all the theoretically (meta-)stable compounds one could possibly realize in this universe. 
To paraphrase Buckingham and Utting, a compound ``{\em ... is a group of atoms ... with a binding energy which is large in comparison with the thermal energy kT}''~\cite{buckingham1970moleculeDefinition}. 
\textcolor{brown}{In other words, with respect to all its spatial degrees of freedom, it is that locally averaged atomic configuration, for which the free energy is in a local minimum surrounded by barriers sufficiently high to prevent spontaneous reactions within some observable life-time.}
As such, CCS depends on external conditions. 
It looses all meaning, for example, when conditions are such that bonding spontaneously emerges and vanishes (aggregation state of plasma). 

The mathematical number of compounds grows explosively with number of constituting atoms due to the mutual enhancement of combinatorial scaling at three rather distinct but well established energetic scales:
First, the number of possible stoichiometries for any given system size (in terms of electrons and total proton number) represents an integer partitioning problem which grows combinatorially, see Ref.~\cite{anatole-ijqc2013} for example. 
The energetic variance among compounds that differ in stoichiometry is on the scale of chemical bonding due to having different number and different types of atoms.  
Second, the number of possible connectivity patterns, i.e.~incomplete labeled undirected weighted graphs distinguishing constitutional isomers/allotropes (commonly drawn as Lewis structures), is mathematically known to grow combinatorially with number of atoms~\cite{Faulon1994,CanonizerMeringer2004,ReymondChemicalUniverse}. 
The energetic variance among constitutional isomers is on the scale of differences in chemical bonding. 
Third, the number of possible conformational degrees of freedom grows combinatorially with number of atoms in a molecular graph (cf.~Levinthal's paradoxon for polymers). And one could even consider different atomic configurations of disconnected graphs, i.e.~macro-molecular or molecular condensed systems, 
to fall into this category of isomers. 
As such, the energetic variance among conformational isomers is on the scale of non-covalent intra- as well as inter-molecular interactions.
We note that stereoisomerism typically occurs among constitutional and conformational isomers. Its extension to compositional chirality has only been proposed only recently~\cite{AlchemicalChirality}.
Seen such size and diversity, highly universal and transferable methods are in dire need, in order to meaningfully explore CCS in search of deepened chemical insight and intuition, and of new compounds and materials which exhibit desirable properties. 
While quantum mechanics and statistical mechanics offer the appropriate physical framework for dealing with CCS in an unbiased and universal manner, the computational complexity of the 
equations involved has hampered their widespread use.

We note that our {\em ab initio} definition of CCS implies that only those compounds are part of CCS that should, at least in principle, be experimentally accessible as long as sufficiently  sophisticated synthetic chemical procedures and reservoirs of the necessary chemical elements are available. While any such synthetic procedure would have to follow the corresponding relevant free energy paths, navigating the virtual analogue of CCS we do enjoy more design freedom and can, namely for any property that is a state-function, also exploit unrealistic fictitious transformations in line with Hess' law, i.e.~without the need for direct correspondence to experimental realization (cf.~`alchemical' transmutations).

We conclude this sub-section by noting that our definition generalizes the more commonly made reference to CCS which typically excludes  conformational isomers, reactive intermediates, or minima in electronically excited states.
For example, first steps towards an {\em ab initio} based representative exploration of the latter were also proposed in 2013 for drug-like compounds by Beratan and co-workers~\cite{BeratanUnchartedCCS2013}.
However, for this review we do not assume the most general view on CCS which would still be consistent with quantum mechanics. Namely that CCS comprises any chemical system, i.e. compounds with {\em any} chemical composition and {\em any} atomic configuration (being close to some state's energy minimum or not). Such an encompassing definition would sacrifice the minimal free energy requirement mentioned above, and it would trivially correspond to the entire domain of CCS. \textcolor{brown}{Therefore, it would forego the useful link to observable lifetimes of systems, as well as the appealing complementarity (not to be confused with orthogonality)} to the well established problem of sampling potential energy hyper surfaces to study free energies or competing elementary reaction steps.

\subsection{Machine learning the potential energy surface}

\textcolor{red}{While QM based studies of CCS are mostly concerned with (meta-)stable compounds,
from inspection of the electronic Hamiltonian it is quite clear that the effect of nuclear charges and 
nuclear coordinates are intimately linked. The well known cusp condition due to Kato's theorem~\cite{kato1957} explicitly links these two variables through the electron density observable. 
As such, ab initio studies of the PES aimed at calculating  geometric distortion, transition states, or
or statistical mechanical averages are closely related to the topic of this review.
More specifically,  early attempts of QML have focused on the PES of homonuclear system (e.g., diamond~\cite{GAP} or Si$_n$ cluster~\cite{BartokGabor_Descriptors2013}) due to its relative simplicity (cf. compositional degree of freedom),
for which many QML methods developed are also applicable to CCS.
The distinction between CCS and the PES is somewhat arbitrary. 
For example, some molecular quantities of significant interest, such as libraries of ensemble properties of protein-ligand binding free energies, require accurate potentials as well as representative sampling of CCS. 
Also, instead of considering (meta-)stable constitutional or conformational isomers as distinct compounds they can also equally well be viewed as local minima of the gloabal PES hypersurface. 
}

\textcolor{red}{As mentioned above,} within studies of the PES, the
focus (at least currently)  is typically placed on a single system, and on computing energies and forces from scratch, i.e.~{\em ab initio}. As such, one does not exploit correlations, constraints, and relationships which only emerge through relationships observed among constitutional and compositional isomers, i.e.~throughout the entire CCS-dimension. 
The most common use-case of quantum methods for atomistic simulations deals with the problem of sampling the configurational degrees of freedom of the atoms of a given system.
\textcolor{red}{To develop a better informed understanding of the field, 
we now also briefly discuss relevant and select machine learning studies
which touch upon the quantum based understanding of CCS but which
primarily are concerned with the PES.
}

The question of how to best 
\textcolor{red}{model a PES using some (physical or surrogate) function approximator and based on} scarce and expensive potential energy surface data sets of specific systems, i.e.~not through CCS, 
obtained from computationally demanding calculations, is long-standing.
Modeling potential energy hyper-surfaces, \textcolor{red}{was} traditionally studied for the purpose of 
molecular spectroscopy or for molecular dynamics applications of a given system. 
\textcolor{red}{The development of empirical interatomic potentials,
particularly the reactive force-field (ReaxFF) approach developed by van Duin and coworkers since 2001~\cite{ReaxFF2001, senftle2016reaxff},
amounts essentially to a traditional multi-dimensional regression problem for fixed functional basis-functions,
and constitutes one of the mainstream efforts into this active field.
Unlike traditional force field approaches, ReaxFF requires no predefined connectivity between atoms (topology), and casts the empirical interatomic potential within a formalism
of bond-order, which depends on the interatomic distances only.
This improved adaptation of an atom to its environment in ReaxFF allows for accurate descriptions of bond breaking and bond formation,
and has been applied extensively to model reactive chemistry at heterogeneous interfaces, involving typically very large systems~\cite{senftle2016reaxff}, made up of millions of atoms.}

\textcolor{red}{The force-field approach, despite its efficiency, its chemical motivation, and its broad applicability and potential accuracy, 
suffers from the fixed functional forms imposed when relying on empirical interatomic potentials, implying that the model is hard to improve by adding more training data,
and could even fail catastrophically in certain regimes and classes. 
This limitation motivates interest in more flexible data-driven fitting model.}
For example, already in 1994, Ischtwan and Collin's \textcolor{red}{improved the Shepard interpolation scheme for PES approximations~\cite{Collins1994TaylorShepardInterpolation}}. 
\textcolor{red}{This illustrates the close relationship to QML: They authors utilised a formalism very similar to the modern kernel ridge regression, one of the work-horses of QML. 
The authors also already} discussed \textcolor{red}{one of the frequent challenges coming along with any new ML model project:}
How to best down-select optimal configurations for minimal data acquisition and training costs, 
and how to obtain systematic model improvements with \textcolor{red}{increasing} training set size~\cite{Collins1994TaylorShepardInterpolation}. 
\textcolor{red}{Early awareness of the trade-off between accuracy and training-cost 
was also already addressed in the 1981 paper by Wagner, Schatz and Bowman:} 
Given a finite compute budget, data for which training instances should be acquired in order to obtain the 
most accurate model of the potential energy hypersurface?\cite{WagnerSchatzBowman_PESfit1981}. 
\textcolor{red}{When facing the exploration of CCS with QML models, analogous questions must be addressed.}
For references to \textcolor{red}{similar studies related to the problem of potential fitting and} preceding 1989, we refer to the comprehensive review by Schatz~\cite{Schatz_PESreview1989}.


\begin{figure}
\includegraphics[scale=3.0]{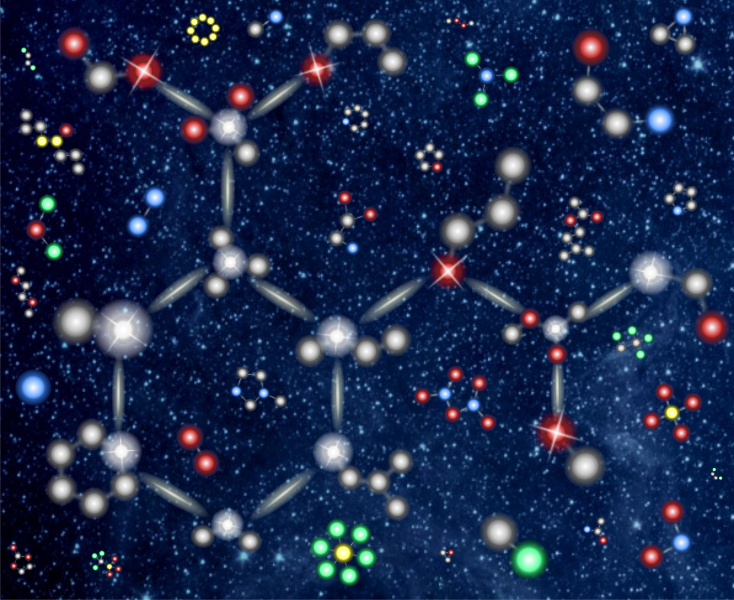}
\caption{\label{fig:Constellation}
A cartoon of similarities among atoms across 
chemical compound space, not in conflict with quantum mechanics.
The exemplary molecule aspirin is highlighted by bonds, and each of its atoms is superimposed with a similar atom in another molecule (hydrogens omitted for clarity).
Green, yellow, gray, red, blue refers to
sulfur, phosphor, carbon, oxygen, and nitrogen, respectively. 
Reproduced with permission from Ref.~\cite{Anatole2020NatureReview}. Copyright 2020 Springer Nature.}
\end{figure}

\textcolor{red}{Most of the data-driven models in the nineties favoured the neural network regressors for PES fitting.}
More specifically, in 1992 Sumpter and Noid published a neural network model for macro-molecules~\cite{SumpterNoidNeuralNetworks1992}.
Additional neural network potentials were published by Blank et al. in 1995~\cite{blank1995neural} \textcolor{red}{for the CO/Ni(111) system},
and Brown et al. in 1996~\cite{brown1996combining} \textcolor{red}{for the study of ground state vibrational properties of two weakly bound molecular complexes: (FH)$_2$ and FH–ClH}.
Neural networks were revisited for systems with increased size the same year by Lorenz, Gross, and Scheffler\cite{Neuralnetworks_Scheffler2004} \textcolor{red}{for H2/K(2$\times$2)/Pd(100) (with substrate fixed)}, followed by
their application to represent high-dimensional potential energy surfaces  \textcolor{red}{for H$_2$O$_2$} by Manzhos and Carrington in 2006~\cite{NN_Tucker2006}.
\textcolor{red}{
Even larger systems, i.e., water clusters (up to 6 untis), were dealt with by Handley and Popelier from 2009 onward~\cite{MachineLearningWaterPotential_Handley2009}, accounting for important electrostatic properties through learning the atomic multipoles.
These early developments used Cartesian/internal coordinates directly as the input of NN models,
which is justified for modeling small to medium-sized systems;
for large systems however, this set-up proves to be too inefficient.} 
In 2007, Behler and Parinello published much improved deep neural network-based potentials~\cite{Neuralnetworks_BehlerParrinello2007}, \textcolor{blue}{encoding molecular geometry effectively in terms of rotation, translation, and permutation invariant atom-centered symmetry function (ACSF)}, 
followed by molecular dynamics applications using meta-dynamics to identify Si phases under high pressure~\cite{Neuralnetworks_Behler2008}.
A detailed overview of various neural network based advances since 2010 was given in 2017~\cite{behler2017first}.
Starting in the same year, multiple, more universal, neural network models were introduced.
In particular, Smith et al. advanced the idea of Behler's symmetry functions in neural networks with the aforementioned normal mode displacements in order to generate a powerful neural network trained on millions of configurations of tens of thousands of organic molecules, called ANI~\cite{ANI_IsayevRoitberg2017}. 
An accurate and transferable neural network exploiting an 'on-the-fly' equilibration of atomic
charges was introduced that same year by Faraji et al.~\cite{Goedecker2017neuralnet}.
And soon thereafter, equally universal neural nets SchNet~\cite{SchNet} and PhysNet~\cite{unke2018reactive} were published.
An extensive review on neural network potentials for modeling the potential energy surfaces of small molecules and reaction is also part of the present issue in Chemical Reviews~\cite{Sergei2020chemicalreview}.

\textcolor{red}{Kernel models started to play a noticeable role for PES fitting in the late nineties}.
In 1996, Ho and Rabitz presented kernel based models for the fitting of potential energies~\cite{Rabitz1996} \textcolor{red}{for three small systems, He-He, He-CO and H$_3^+$}.
\textcolor{red}{Similar to early PES fitting works within NN framework, these early kernel-based models also utilised simple cartesian/internal coordinates as input,
and therefore applicability was limited.}
While the mathematics of kernel-based surrogate models was firmly established many decades ago, \textcolor{red}{only from 2010 and onward, kernel-based models began to flourish building on the seminal work contributed by} Csanyi, Bartok and co-workers within their 'Gaussian-Approximated Potential' (GAP) method, relying on Gaussian process regression (GPR) and an atom index invariant bispectrum representation\cite{bpkc2010}.
In 2012, Henkelmann and co-workers introduced an interesting application of support vector machines (SVM) towards the identification of transition states~\cite{ML4Graeme2012}.
One year later, the first flavor of the smooth overlap of atomic positions (SOAP) representation for GPR based potentials was published~\cite{BartokGabor_Descriptors2013}.
The SNAP~\cite{SNAP_Aidan2015} method popularized the GAP idea using linear kernels in 2015, as well as other GPR applications with automatically improving forces were published the same year~\cite{Zhenwei2015,RampiMLQMMM}.
Around the same time, a first stepping stone toward a universal force-field, trained on atomic 
forces throughout the chemical space of molecules displaced 
along their normal modes, was established~\cite{MLatoms_2015}.
Reproducing kernels were also shown in 2015 to be applicable towards dynamic processes
in biomolecular simulations~\cite{soloviov2015reproducing}.
And ever more accurate GPR based potentials were introduced in 2016~\cite{CovariantKernelsSandro2016},
and in 2017~\cite{chmiela2017machine,unke2017toolkit}.
GPR was also applied to challenging processes in ferromagnetic iron~\cite{GaborMarzari2018GAPandIron} and
to the problem of the on-the-fly prediction of parameters in
intermolecular force-fields~\cite{bereau2018non}.
Amorphous carbon was studied using SOAP based GPR/KRR models~\cite{deringer2018computational,caro2018reactivity}, 
and GDML, another series of highly robust and accurate GPR/KRR based molecular force-fields, was introduced in Refs.~\cite{chmiela2017machine,chmiela2018towards,chmiela2019sgdml,sauceda2020GDML} starting in 2017.

\textcolor{red}{GPR and NN are currently the two most popular regressors for PES fitting and each exhibits advantages and disadvantages.
Seemingly very different in design, they do resemble each other to some extent in the sense that they take the role
of basis-functions (to be elaborated in section `Regressor'),
though the similarity may be blurred within the framework of deep NN.
Numerical comparison of the performance of these two methods is interesting.
Most notably, such a comparison was made}
for modeling the potential energy surface of formaldehyde 
in 2018 by Manzhos and co-workers~\cite{Sergei2018kernelsVSneuralnets}.
A similar yet independent study on the same system was performed in 2020 by Meuwly and co-workers~\cite{silvan2020formaldehyde}. 
Both studies confirm that kernel based QML models reach higher predictive power than neural network based models for same training set sizes.
\textcolor{red}{A highly related comparative study on} modeling
vibrations in formaldehyde was contributed by K\"aser et al.~\cite{silvan2020formaldehyde} also in 2020.

\textcolor{red}{As for the active learning of interatomic potentials, most of the related studies relied on the kernel framework, some of them also detailed below in the section `Training set selection'. As early as in 2004, }De Vita and co-workers proposed updating potential parameters to ab initio results during molecular dynamics runs ('learning-on-the-fly') \cite{lotf2004}\textcolor{red}{, for a very large system, i.e., silicon systems composed of up to $\sim$200,000 atoms, though the reference level of theory is quite approximate.}
\textcolor{red}{Podryabinkin and Shapeev proposed the so-called $D$-optimality criterion for selecting the most representative atomistic configurations for training on-the-fly
in as early as 2017~\cite{Shapeev_AL}}.
Using kernel ridge regression (KRR, a variant of GPR), Hammer and co-workers revisited the on-the-fly-learning idea for structural relaxation in 2018~\cite{jacobsen2018fly}, and investigated the exploration vs.~exploitation trade-off~\cite{jorgensen2018exploration}.
In 2019, Weinan, Car and co-workers contributed another active learning procedure for accurate potentials of Al-Mg alloys~\cite{RobertoCarWeinan2019activelearning}, and
Westermayr et al.~extended the use of neural networks for molecular dynamics in the electronic ground-state towards photo-dynamics~\cite{westermayr2019machine}.
\textcolor{red}{Among the many purposes (also challenges) of QML for PES, one particular is to scale to extremely large (thus more realistic) system.
Numerous efforts have pushed us closer to this goal, and most notably,}
Weinan, Car and co-workers made full use of the Summit supercomputer to simulate 100 
million atoms with {\em ab initio} accuracy using convolutional 
neural networks~\cite{RobertoCarWeinan2020pushing100m}, for which they they subsequently 
were awarded the Gordon Bell prize 2020 by the {\em Association for Computing Machinery}.

\subsection{Navigating CCS from first principles}
\label{sec:FPCCS}


The scientific research question of how properties trends across CCS lies at the core of the chemical sciences. 
Due to ever improving hardware performance, 
improved approximations to Schr\"odinger's equation, 
most notably within density functional theory and localized
coupled cluster theory, QM data-sets of considerable size have emerged, enabling the use of statistical learning to train surrogate QML  models which can provide accurate and rapid quantum property estimates for new compounds within their applicability domain.

While quantum mechanics based computational materials design efforts had been undertaken as early as the nineties~\cite{Beratan1991,Beratan1996,ceder1998predicting,ZungerNature1999} with important progress made in the eighties~\cite{NorskovPRL2002,anatole-prl2005}, the first principles based computational high-throughput design has by now become an important success story~\cite{Marzari20202highthroughput2Dmaterials}.
First attempts to employ machine learning and quantum predictions to discover new ternary materials data bases date back to seminal work by Hautier and Ceder in 2010.~\cite{MachineLearningHautierCeder2010,Hautier2020ChemistVsMachine}.


\textcolor{red}{As a promising alternative of ab-initio high-throughput computations (or solving Schrodinger equation in general),
QML assumes locality of atom in molecule when constructing the mapping from molecular distance/similarity to difference in properties within QML,
and the final predictive performance depends on how similar two local (and thus global) entities are, i.e., nuclear types covered by test set are required to be retained in the training set.
The capability of QML to treat species made up of elements not seen in training set is however very limited, if not impossible at all.
Luckily, there exists the so-called ``alchemical'' methods, being quite different in philosophy, 
allowing for effective and efficient treatment of the change of nuclear types, with or without the constraint that the number of electron number ($N_e$) being fixed (i.e., isoelectronic).
We note by passing that alchemy is always established within the density functional theory framework,
as it would be tremendously simpler to expand molecular property (mostly energy) as a function of four variables ($x$, $y$, $z$ and $Z$) than $3N$ ones in the case of wavefunction-based formulation.
Furthermore, it could be well fit into the inverse design problem due to its analytical nature.}

Previous methodological works tackling chemical compound space from first principles
through variable (`alchemical') nuclear charges \textcolor{red}{are contributed by many pioneers, }including Wilson's formal 
4-dimensional density functional theory~\cite{WilsonsDFT} in 1962, \textcolor{red}{which expresses the exact non-relativistic ground state energy of an electronic system as a functional of the electron density, which per se is a function of the spacial coordinates and nuclear charges}.
\textcolor{red}{Following Wilson's idea, Politzer and Parr~\cite{PolitzerParr} in 1972 made one step further towards practical computation by transforming Wilson's formula into functional of the total electrostatic potential $V(r,Z)$
and based on approximations of which derived some useful semi-empirical formulas for the total energy of atom and molecule, through} the use of thermodynamic integration~\cite{PolitzerParr}.
\textcolor{red}{Later in the eighties, Mezey made some interesting discoveries~\cite{Mezey1980,Mezey1986},
about the global electronic energy bounds for a variety of isoelectronic polyatomic system,
which may be found useful for quantum-chemical synthesis planning, using multi-dimensional potential surfaces.}

\textcolor{red}{The theoretical alchemical research was resurrected in the new millennium. Among the numerous contributions, notable ones include a}
variational particle number (variable proton and electron number) approach for rational compound design~\cite{anatole-prl2005} \textcolor{red}{proposed by one of the authors and collaborators},
\textcolor{red}{followed by a more detailed description of the underlying theories, in the name of}
molecular grand-canonical ensembles (GCE) ~\cite{anatole-jcp2006-2}.
\textcolor{red}{In the same year,}
a reformulation \textcolor{red}{of GCE} in terms of linear combinations of atomic potentials (LCAP)~\cite{RCD_Yang2006} \textcolor{red}{(instead of $Z$ and $N_e$ as in GCE)} \textcolor{red}{was proposed by Wang, et al.,
but for the optimization of molecular electronic polarizability and hyper-polarizability,
with the optimal molecule determined analytically
in the space of electron-nuclei attraction potentials.}
\textcolor{red}{For the isoelectronic case, related works include the development of ab-initio methods for the computation of} higher order alchemical derivatives~\cite{LesiukHigherOrderAlchemy2012} \textcolor{red}{by Lesiuk et al. in 2012},
\textcolor{red}{as well as the assessment of} the predictability of alchemical derivatives~\cite{munoz2017predictive} \textcolor{red}{by Munoz et al. in 2017}.
\textcolor{red}{More recently,} alchemical normal modes in CCS~\cite{AlchemicalNormalModes},
alchemical perturbation density functional theory~\cite{APDFT}, 
and even a quantum computing algorithm for alchemical materials optimization~\cite{barkoutsos2020quantum} \textcolor{red}{were proposed, further enriching the field}.

Starting in 1996 with stability of solid solutions~\cite{Marzari_prl_1994},
multiple promising applications, based on quantum alchemical changes, 
have been published over recent years, including
thermodynamic integrations~\cite{ArianaAlchemy2006},
mixtures in metal clusters~\cite{AlchemicalDerivativeBinaryMetalCluster_WeigendSchrodtAhlrichs2004,weigend2014extending}, 
optimization of hyperpolarizabilities~\cite{MolecularDesignRinderspacherBeratanYang2009},
reactivity estimates~\cite{CatalystSheppard2010},
chemical space exploration~\cite{CCSexploration_balawender2013},
covalent binding~\cite{Samuel-JCP2016},
water adsorption on BN-doped graphene~\cite{al2017exploring},
the nearsightedness of electronic matter~\cite{StijnPNAS2017},
BN-doping in fullerenes~\cite{balawender2018exploring},
energy and density decompositioning~\cite{AtomicAPDFT}, 
catalyst design~\cite{Keith2017alchemicalCatalysis,Keith2018benchmarkingalchemy,Keith2020acceleratedCatalystAlchemy},
and 
protonation energy predictions~\cite{von2020rapidDeprotonation,munoz2020predicting}
Symmetry relations among perturbing Hamiltonians have also enabled the discovery of `alchemical chirality'~\cite{AlchemicalChirality}. 

An extension of computational alchemy towards descriptions which go beyond the Born-Oppenheimer approximation has been introduced within path-integral molecular dynamics, enabling the calculation of kinetic isotope effects, already in 2011~\cite{alejandro-jctc2011}, and subsequently by Ceriotti and Markland~\cite{ceriotti2013efficient}. 

But also varying the electron number is a long-standing concept within conceptual DFT~\cite{Geerlings_DFTConcepts,FracN_YangAyers}. 
Actual variations have only more recently been considered, e.g.~to estimate redox potentials~\cite{anatole-jcp2006-2,RedoxPotentialQMMMalchemy}, higher order derivatives~\cite{LesiukHigherOrderAlchemy2012,munoz2017predictive,AlchemicalNormalModes},
or for the development of improved exchange-correlation potentials~\cite{DiscontinuousXC_MoriSanchezCohenYang2009}.

\section{Heuristic approaches}


Modern systematic attempts to establish quantitative 
structure-property relationships (QSPRs) have led to 
computationally advanced 
bio-, chem-, and materials-informatics methodologies.
Unfortunately, conventional approaches in QSPR predominantly rely on heuristic assumptions about the nature of the forward problem, and are thus inherently limited 
to certain applicability domains. 
The implicit bias, often due to lacking basis in the underlying physics is known, as discussed e.g.~in a 2010 review by G.~Schneider~\cite{SchneiderReview2010}, 
and many great improvements have been contributed more recently~\cite{AspuruOpreaRoitbergTetko2020qsar}.

While heuristic in nature, QSPR can still provide useful qualitative trends and insights for relevant applications, and sometimes yield accurate predictions for specific property sub-domains and systems.
Albeit not directly relying on the laws of quantum mechanics, these early developments are still valuable,
\textcolor{red}{in the sense that some just correspond to special variants of the more complicated models, for instance, linear model can be mapped onto the framework of kernel method,
by choosing a linear kernel, instead of say Gaussian kernel for Gaussian process regression (GPR).}
\textcolor{red}{Other heuristic approaches, exhibiting more quantitative characteristics,} can be considered important precursors for modern QML.
\textcolor{red}{Such examples include Collin's improved Shepard interpolation scheme~\cite{Collins1994TaylorShepardInterpolation} 
for accurate representation of molecular potential energy surfaces,
which resembles the form of kernel methods except that
the weights are determined in a heuristic way, instead of being regressed as in GPR.
One may also argue that Collin's scheme could be recast into the kernel framework, except that
a specific kernel is chosen such that the Shepard interpolation weights in Collin's scheme are exactly reproduced (with the constraint that these weights sum up to 1).
Another highly related concept is Ramon Carbo-Dorca's quantum similarity (for a comprehensive review, see~\cite{bultinck2005QuantumSim}),
derived based on density matrix, or molecular orbitals, or other related quantum quantity,
is also closely linked to kernel based methods,
and may be used directly as parameter-free kernel matrix elements
(unlike in GPR, kernel matrix element characterizing similarity is typically hyper-parameter dependent).
}

In subsections below, we focus on relevant literature regarding three distinct perspectives which largely follow chronological order:
i) low-dimensional correlations or simple models from the early days of chemistry;
ii) coarse representations of molecules and derived quantities, mostly providing an overview of QSPR;
iii) molecular representations based on properties.

\subsection{Low-dimensional correlations}\label{sec:simple_model}
Early practices of fundamental chemical research dealt with spotting correlations between inherent properties of the system and systematic changes of observed quantities.
Possibly the most famous example for such work is Mendeleev's discovery of the periodic table~\cite{mendeleevs_PT}. 
Other important examples correspond to
Pauling's electronegativity concept and covalent bond 
postulate~\cite{pauling1932CovalentBondsPNAS},
or Pettifor's Mendeleev numbering scheme~\cite{pettifor_scale, modified_pettifor_scale}.
Work along such lines has been continued, and recent contributions include revisiting Pettifor 
scales~\cite{Gross2016periodictable,Oganov2020nonempiricalCCS},  use of variational autoencoders to `rediscover'
the ordering of elements in the periodic table~\cite{glushkovsky2020periodictable},
or the chemplitude model which extends Pauling's concept~\cite{chemplitude}, among many others. 
Free-energy relationships are the subject of yet another broad category of early research which is still active today. 
Relating logarithms of reaction constants
(free energy difference) across CCS for related series of reactions 
~\cite{linear_dG_relation}, has led to the famous Hammett equation, a 2D projection of all degrees freedom onto composition and reaction conditions~\cite{hammett_equation, hansch_survey_Hammett_relation_1991, hammett_relations_1935cr}. Similarly low-dimensional
effective degrees of freedom have been identified within  Hammond's postulate\cite{hammond_correlation}, or Bell-Evans-Polanyi principle~\cite{BEP_1, BEP_2}.

Most of the aforementioned concepts were proposed to gain a better (or more useful) understanding of molecular behavior in the first place. For extended systems such as metallic surfaces, complexities arise and many of the simplified molecular models are no longer applicable. 
With the advent of density functional theory (DFT)~\cite{KS, HK,pbe,pbe01,PBE02}, alternative descriptors have been proposed during the past decades, playing an increasingly important role. 
Notable contributions include the $d$-band center model by Hammer et al.~\cite{hammer_electronic_1995, hammer_noblity_Au_1995},
the generalized coordination number~\cite{vallejo_GeneralizedCN_theory, vallejo_GeneralizedCN}, or the Fermi softness~\cite{Fermi_softness}.
Free-energy relationships are more robust against subtle changes in the electronic structure and are being widely applied in analyzing surface elementary reaction steps~\cite{van_santen_BEP_TMsurf_2010}.
Scaling relations between the energetics of adsorbed species on surfaces~\cite{scaling_relation_Eads_on_metal_surf}, also enjoy extensive attention, and have been proven useful for catalyst design, regardless of the surface being metallic not.~\cite{scaling_relation_Eads_on_oxide, scaling_Eads_origin,ReviewCatalystNorskov}
Many of the empirical chemical concepts such as electronegativity, softness/hardness, electrophilicity/nucleophilicity can be rationalized and quantified within what is known as `conceptual' DFT~\cite{Yang_dft_book, geerlings_cdft_2003}. 
This specific field, as pioneered by Fukui or Parr and Yang~\cite{Yang_dft_book}, has been 
championed and furthered by many including Geerlings, De Proft, Ayers, Cardenas, and co-workers~\cite{geerlings_cdft_2003, geerlings_cdft_linear_response_2014}. 

We note that simple models, involving one or few variables in general, represent effective coarse-grained schemes applicable to specific sub-domains of chemistry. 
While they lack the desired transferability of quantum mechanics they often do encode well tempered approximations, and therefore are capable of capturing much of the essential physics. As such, they have much to offer, and they could, for example, serve the design of robust and general representations enabling the training and application of improved QML models (see below or Refs.~\cite{BAML, MachineLearningMeetsQuantumPhysics2020book}). 
Alas, this idea, to connect traditional QSPR, based on well established heuristics, with more recently developed generic ML models, is still largely unexplored, despite the fact that the latter often bear (magic) black box characteristics allowing for little qualitative insights. 
Unifying modern ML with traditional QSPR could therefore also help resolve open challenges in QML. For instance, how can we properly represent different electronic (spin-) states of molecules in the molecular representation, or different oxidation states? Conceptual DFT derived linear or quadratic energy relationships suggest treating number of electrons ($N_e$) and/or its powers as independent features might be a reasonable starting point.
Another thus-inspired direction of research is to utilize conceptual DFT-based local indicators as properties of  composing atoms/bonds/fragments of a target molecule as a starting point (much like the fundamental variables such as $Z$ and $R$) for building representations. This might be necessary in order to address hard and outstanding problems such as building QML models of intensive properties or to account for the multi-reference nature.

\subsection{Stoichiometry} \label{sec:stoichiometry}

Given a fixed pattern of structure, stoichiometry alone  can be used as a unique representation of the system under study.  
This idea has been demonstrated for an 
exhaustive QML based scan of the elpasolite (ABC$_2$D$_6$ stoichiometry) sub-space of CCS, 
predicting cohesive energies of all the 2 million  crystals made up from main-group elements~\cite{Elpasolite_2016}.
Elpasolites are the most abundant quaternary crystal form found in the Inorganic Crystal Structure Database.
Comparison of the QML results to known competing ternary and binary phases enabled favourable stability predictions for nearly 90 crystals (convex hull) which subsequently have been added to the Materials Project data-base~\cite{MaterialsProject}. 
A compact 
stoichiometry based representation in terms of 
period and group entry for elements A, B, C, and D
was shown to reach explicit geometry based many-body potential representations at larger training set size, indicating the dominance of the former in large training data regimes~\cite{FCHL}. 
Similar work was subsequently done by Ye et al.~\cite{ye2018_DNNsolidStablity}, as well as Marques et al. for perovskites 
on crystal stablity~\cite{schmidt2017predicting}, as well as by Legrain 
et al.~\cite{Legrain2017_formulaML4ICSD} for predicting vibrational free energies 
and entropies for compounds, drawn from the Inorganic Crystal Structure Database.

A naive but useful derived concept is the so called ``dressed atom'' concept~\cite{BoB}, which characterizes the atom in a molecule of a fixed stoichiometry. 
When using this approach together with a linear regression model to approximate the total energy (or atomization energy), the accuracy turns out to be surprisingly reasonable~\cite{QM9}, at least for common data sets of organic molecules with small variance among constitutional isomers.
For instance, the corresponding mean absolute error (MAE) for QM7 dataset is \emph{only} 15.1 kcal/mol~\cite{BoB}.
Using bond counting, the MAE could be improved further to less than 10.0 kcal/mol, within reach of a conventional DFT GGA functionals~\cite{ChemistsGuidetoDFT}.
Therefore, it seems advisable to always use the dressed atom approach for centering the data for any fixed stoichiometry (i.e.~averaging out constitutional and conformational isomers) before proceeding to the next level of QML training on the complete set of degrees of freedom. 
This normalization step can also be seen as data preprocessing, enabling the QML model to focus on ``minor'' system-specific deviations from the mean~\cite{Elpasolite_2016,Amons}.

\subsection{Connectivity graph} 
When the systems under study do not share some common structural skeleton, stoichiometry alone is not enough, and the covalent bonding connectivity between atoms, 
as well as conformations, may have to be also examined for a general description of the systems.

It is worth pointing out that chemists often assume a one-to-one relationship between the molecular graph and its associated global conformational minima 
(or the second lowest energy minima, or the third, etc).  And therefore it should be possible to build a QML model to predict relevant quantum properties of such ordered minima 
from graph-input only.
In fact, the remarkable performance of extended H\"uckel theory for some systems could be explained in this way. 

Due to the intuitive accessibility and applicability of (incomplete) graph based representations, such as Lewis structures and their extensions, for a wide range of molecular systems (mostly bio- and or organic systems), associated ML methods have received broad attention and wide applications in many fields such as cheminformatics or bioinformatics.
Examples of such representations include various fingerprint representations~\cite{DescriptroOverviewMeringer2005}, such as the signature methodology~\cite{Visco2002,SignatureFaulon2003,FaulonProtein2005}. 
Another notable example corresponds to the so-called extended circular fingerprint (ECFP)~\cite{ecfp}. 
    ECFP and similar representations have been used for drug design~\cite{ecfp_drug_activity,  ecfp_ligand_design} and qualitative exploration of CCS~\cite{GDB17_166B,gdb17_exploration}. 
ECFP has also been used in KRR models for prediction of quantum properties of QM9 molecules. 
Numerical results for ECFP based QML models indicate a a substantially worse performance compared to more complete, geometry derived representations~\cite{googlePaper2017}.

Modern molecular graphs, typically in SMILES format, based neural networks models, have gained considerable momentum during the past decade. A vast amount of related literature deal with chemical synthesis and retrosynthesis in such representation spaces (mainly in organic chemistry)~\cite{schwaller_found_2018,schwaller_molecular_2019} typicaly favoring different deep learning architectures,
chemical reaction network~\cite{Reiher2020_reaction_network},
as well as molecular design using variational autoencoder (VAE, which maps a molecule represented by SMILES string to some latent space~\cite{chemical_design_VAE}), or absence or presence of relationships between functional groups and binding affinity as also recently 
explored through use of random matrix theory in drug design~\cite{yang2019ligand}.
 The incorporation of new and improved formats, such as SELFIES~\cite{SELFIES}, might still lead to further improvements for such research. 

In the context of a first principles view on CCS, we note however that molecular graphs only encode a (biased) statistical average of the many conformational configurations for a molecule near some local minima in the potential energy surface. As such, they are naturally disposed for use of QML models of ensemble properties. Work along such lines still awaits to be explored in the future.
Albeit popular and justified for certain problems, graph-based approaches are inherently limited 
when it comes to non-covalent problems, such as supra-molecular assembly processes governed by van der Waals interactions, 
metal cluster/bulk/surface adsorption involving ``multivalent'' (transition) metal elements controlled 
by weaker metallic bond (c.f., covalent bond), 
or chemical reactions requiring the transformation of graphs from one into the other. 
In such situations, the intuitive concept of a graph is  ill-defined, and the necessary corrections are not always obvious.

\subsection{Coarse-grained} 
\label{sec:cg}
As the system size grows, the cost in training and prediction of QML models increases accordingly, though with more favourable scaling
than typical quantum chemistry methods.
Therefore, it may become very demanding or even impossible to deal with system sizes which cross certain thresholds. 
In such scenarios, one typically represents the systems in a coarse-grained fashion, meaning ``superatom'' (groups of atoms in close proximity, or beads) in a molecule are being considered. 
Coarse-grained approaches can drastically reduce the number of degrees of freedom and are therefore the only feasible way to model systems at macroscopic scale.
More importantly, they enables a significant collapse of
the size of chemical space due to the transferability  of beads by design~\cite{Bereau2019MachineLearningCoarseGrained}.


Current practises of coarse-grained idea comprise mostly coarse-grained force fields (CGFF) for simulation of large systems, such as macro-molecular systems and soft matter.
With the emerging need for systematic control of the accuracy of models of such systems, coarse-grained representation based QML models (CGQML) may be a rather promising alternative to CGFF,
much the same as how QML models based on full-atom representation remedy the deficiencies of classical force field approaches for small to medium-sized molecules~\cite{unke2020MLforFFs}.
Such comparison between QML and FF makes sense, as the most modern implementations of ML hold promise to approach the computational efficiency of FFs. 
Some of the first studies on coarse-grained representation used together with QML include
\textcolor{red}{John and Csányi's free energy surface modelling of molecular liquids in 2017~\cite{john_many_body_2017}. Later, efforts to tackle}
complicated bio-systems
was reported by Bereau and co-workers in 2019~\cite{Bereau2019MachineLearningCoarseGrained}, as well as by Clementi and co-workers~\cite{CGMD2019}.
Compared to CGFF, CGQML could be significantly more accurate once the system information is properly encoded in the coarse-grained representation,
as the QML part can recover what is missing in the CG part by careful selection of training data ({\em vide infra}).

\subsection{Property based}\label{sec:mpx}
There exists another type of representation, typically referred to as descriptor and the least ``ab-initio'' in spirit, in which the basic idea is to simply select a set of pertinent 
atomic/molecular properties as underlying degrees of freedom. The properties can stem either from calculation and/or 
experiment and have to be relatively easy to obtain and are typically supposed to somehow `describe' the property of interest, and hence the name `descriptor'.
This representation is often utilized in combination with some non-linear regressor like neural network, 
as the relationship between the chosen properties is commonly highly non-linear.
Though this approach could be universally applicable, no matter the size or 
composition of target systems, its predictive power is limited by construction due
to its potential lack of uniqueness~\cite{BAML, anatole-ijqc2013}.
Most of the studies following this direction can be traced back 
to the early applications of ML in chemistry and related fields, 
one example being Karthikeyan et al.'s work~\cite{karthikeyan_general_2005} 
on melting and boiling points prediction of molecular crystals using the properties 
of standalone molecules as feature vector.
A more recent, and systematic study of this idea has applied optimization algorithms towards the down-selection of descriptor candidates in order to build predictive ML models of formation energies of binary solids~\cite{ghiringhelli2015big}. 
From a first principles point of view, however, such representations are questionable since relationships between different observables (or other arbitrary mathematical properties), obtained as expectation values of independent operators, are not necessarily well defined.

\section{QML Methodology}
The fundamental idea to employ machine learning models in order to {\em infer} solutions to Schr\"odinger's equation throughout CCS---rather than solving them numerically---was first introduced in 2011~\cite{CMarxiv,CM}. The authors stated
that  {\em "... the external potential ... uniquely determines the Hamiltonian $\rm{H}$  of any system, and thereby the ground state’s potential energy by optimizing $\Psi$"}, and they show that one can use QML instead (encoding the number of electrons implicitly by imposing charge neutrality).
As such, the problem of predicting quantum properties throughout CCS belongs to what is commonly known as `supervised learning'.
One typically distinguishes between unsupervised (compound data only) and supervised (data records including compounds and associated properties) learning. In this review, we focus on the latter, i.e.~on the question how, given sufficient exemplary structure-property pairs, properties can be inferred for new, out-of-sample compounds.

The generic procedure for supervised learning requires to first define the model architecture, i.e.~the mathematical expression for the statistical surrogate model $f$ which estimates some quantum property $p$ as a function of any query compound ${\bf M}$, 
\bea 
p^{\rm{QML}}(\mathbf{M}) &\approx & f(\mathbf{M}|\{\mathbf{M}_i\}, \{ p^{\rm ref}_i\}; \{c_i\})
\eea
where $f$ corresponds to the regressor, and regression coefficients and hyper parameters $\{c_i\}$ are to be obtained via minimization of training loss-function quantifying the deviation of $p^{\rm QML}$ from $\{p^{\rm ref}_i\}$ for all training compounds $\{\mathbf{M}_i\}$. 
In other words, $f$ is parametric in regression coefficients and hyper-parameter which, in return, are non-linear functions in the training data. 
The origin (calculated or measured) as well as the actual existence 
(some properties, such as energies of atoms in molecules~\cite{AtomicAPDFT}, are not observables but can still be inferred) of $p^{\rm ref}$ is secondary. 
Noise in the data (due to experimental or numerical uncertainty, or due to minor inconsistencies) can be accommodated to a certain degree through well established regularization procedures.
Converged cross-validation protocols also help to avoid overfitting and to enable the optimization of hyper-parameters as well as meaningful estimates for any interpolative query. 
\textcolor{red}{For introductory texts on kernel-based regressors, the reader is referred to the book by Rasmussen, et al.~\cite{RasmussenWilliams};
As for representation and training set, several reviews have recently been published~\cite{RaghusReview2016,huang2018quantum,schmidt_recent_2019,unke2020MLforFFs}.}

\subsection{Regressor} \label{sec:intro_regressor}
\begin{figure}
\includegraphics[scale=0.2]{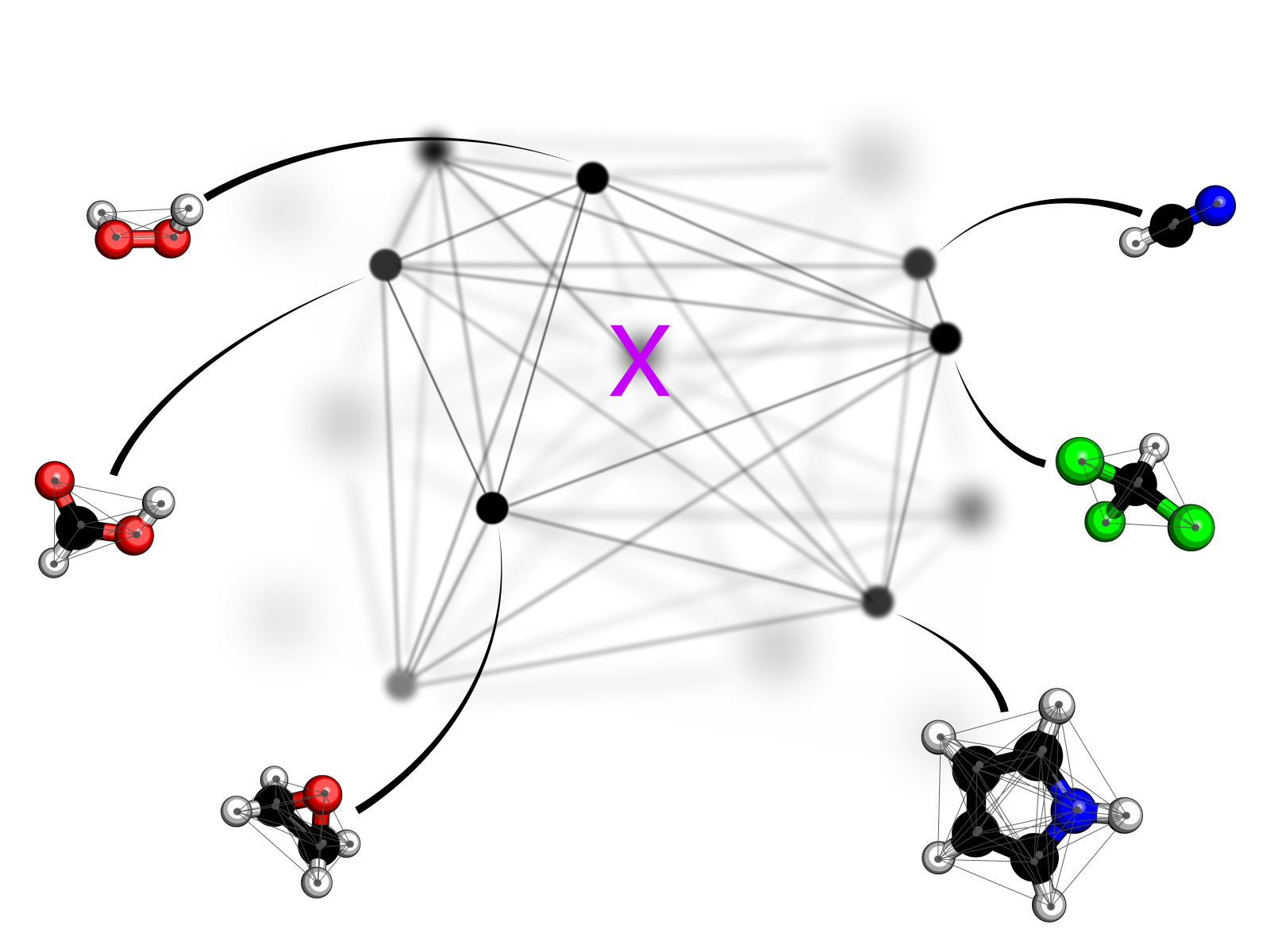}
\caption{\label{fig:kernel}
3D projection of high-dimensional kernel representation of chemical compound space. 
Within kernel ridge regression, chemical compound space corresponds to a complete graph where every compound is represented by a black vertex and black lines correspond to the edges which quantify similarities. 
Each compound, in return, can be represented by a molecular complete graph (e.g.~the Coulomb matrix (CM)~\cite{CM}) recording the elemental type of each atom and its distances to all other atoms. 
Given known training data for all compounds shown, a property prediction can be made for any query compound as illustrated by {\color{pink} \bf X}.
Choice of kernel-function, metric, and representation will strongly impact the specific shape of this space, and thereby the learning efficiency of the resulting QML model.  
}
\end{figure}

When considering the problem of fitting a generic set of basis-functions to pre-calculated data some of the most commonly made choices in the field of atomistic simulation include 
support vector machines (kernel ridge regression) --- tantamount to Gaussian process regression in their specific model form, 
neural networks, 
random forests, 
or permutationally invariant polynomials (PIPs)~\cite{vapnik2013book,RasmussenWilliams,PotentialEnergyFit_Bowman2003}. 
While agnostic about the training labels by construction, the choice of these basis set expansions constitutes a crucial step. 
Most evidently for support vector machines, non-linear kernel functions (based on feature representations {\em vide infra}) 
map any non-linear high-dimensional regression problem 
into a low-dimensional kernel space within which the regression problem becomes linear, and therefore straight-forward to solve through a closed-form expression  (`kernel-trick'). 
How kernel space relates to CCS is also quite intuitive to grasp when thinking about it as a graph of compounds.
\textcolor{red}{As displayed in} Fig.~\ref{fig:kernel},
\textcolor{red}{each compound, being representable by a molecular graph (or derived matrix such as a Coulomb matrix or Cartesian coordinate and nuclear charge vector)  is projected into higher dimensional feature space (shown are only three principal 
dimensions from the infinite number of dimensions defined within the framework of KRR/GPR).
The complete connection between all compounds in the new space form another type of graph,
with each vertex corresponding to a compound,
and each edge corresponding to a similarity measure of compounds (edge length may indicates a metric distance between two compounds).
Inferring the property of a new compound (labeled as pink `X' in Fig.~\ref{fig:kernel}) may be conceptualized as summarizing the weighted properties,
with weights being some function of the lengths of edges formed between X and all other vertexes, as well as among all vertices other than X.
Within this picture it becomes intuitively obvious that the interpolating accuracy must improve with increasing compound density.
}

While deep neural network models are very powerful and possess significant black-box character,
their training requires data sets of very large size as well as substantial
calculation effort in order to optimize the regression coefficients and hyper-parameters (no closed-form solution is known).
In this sense, kernel methods are rather light-weight and preferable in scarce data scenarios, 
as they enjoy the potential benefits of being more intuitive and faster to train.
The specific architecture of the neural network will affect its performance and data-efficiency dramatically. 
Deep, recurrent, convolutional, message passing, generative, adversarial, geometric neural networks, and other flavours, 
as well as choices of activation function, number of layers and neurons, have all shown 
significant impact on the cost of training and on the predictive power in atomistic simulation. 

In the case of GPR/KRR the architecture is much simpler and hence of a lesser concern 
(GPR/KRR can be seen as single layer neural network model in the limit of infinite width~\cite{neal1996_NNasGP}),
but the specific kernel space does not only depend on choice of kernel function but also on choice of metric. 
While it is clear that one should avoid similarity measures which do not meet the mathematical criteria of how a metric is defined (identity, symmetry, triangle inequality),
the impact of the specific metric choice has not yet been studied much in the field of atomistic simulation. 
Euclidean, Manhattan, or Frobenius norms are commonly used. 
Only most recently, the use of the Wasserstein norm has been proposed to gain permutational atom-index invariance while using 
index-dependent matrix representations~\cite{onur2020wasserstein}.
From inspection of Fig.~\ref{fig:kernel}, it should be obvious that any non-linear change in metric will strongly 
affect the shape of the kernel regression space, and thereby the overall performance.

\begin{figure}
\includegraphics[scale=0.45]{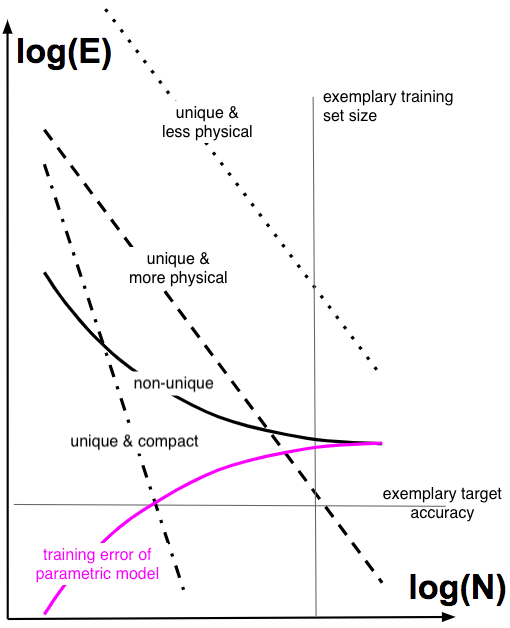}
\caption{\label{fig:LearningCurves}
Illustration of learning curves: Errors (E) versus training set size ($N$).
Horizontal and vertical thin lines illustrate exemplary target accuracy and available training set size, respectively. 
For functional ML models, training errors are close to zero (not shown) and prediction errors must decay linearly with N on log-log scales. Black solid, dotted, dashed and dotted-dashed lines exemplify prediction errors of ML models with incomplete information (ceases to learn for large N due to being parametric, using non-unique representations, or training on noisy data), unique and less physical representation, unique and more physical representation, and explicit account of lowered effective dimensionality (i.e., `compact'), respectively.
The solid pink line corresponds to the training error for a parametric model.
Training errors for ML models are negligible for noise-free data.
}
\end{figure}

\subsection{Learning curves}
Correct implementations of ML algorithms applied to noise-free data sets afford interpolating ML models,
which avoid overfitting and enable statistically meaningful predictions of properties 
of out-of-sample compounds~\cite{vapnik2013book},
after proper
regularization and hyper-parameterization through converged
cross-validation protocols, as discussed in great detail in the literature, for example, in Refs.~\cite{AssessmentMLJCTC2013,RuppTutorial2015}. 

\textcolor{red}{Based on statistical learning theory,} the leading order term of the out-of-sample prediction error (E)
was shown to decay inversely with training set size $N$, i.e.,
E $\propto a/N^b$, for GPR/KRR as well as for neural network models~\cite{vapnik1994learningcurves,StatError_Muller1996}.
\textcolor{red}{This is not surprising,
considering the great similarities shared between NN models and GPR/KRR model,
as is also explained in the preceding subsection.
This asymptotic behaviour for QML models has been confirmed within numerous and independent studies, many of which referenced here within.}
As illustrated in Fig.~\ref{fig:LearningCurves},
learning curves (LC), i.e.~prediction error E vs.~training set size $N$, plotted on log-log scales 
assume linear form ($\log{\rm E} = \log{a}- b \log{N}$),
and serve as a useful standard, facilitating   
systematic comparison and quality control of the efficiency of differing ML models. 
For maximal consistency, the QML models should be trained and tested on the exact same
cross-validation splits stemming from the exact same data set. 
\textcolor{red}{When the data contains noise, or when relevant degrees of freedom are neglected (e.g.~through use of a non-unique representation,
such as the bag of bond (BoB) representation~\cite{BoB}, see subsection~\ref{sec:x_descrete} for more details),
the learning will cease eventually for some training-set size, manifesting itself visually through learning curves which level off, cf.~solid black line in Fig.~\ref{fig:LearningCurves}. 
For noise-free data and complete representations, however,  a linear correlation between log(E) and log(N) is to be expected (see the dotted and dashed lines in Fig~\ref{fig:LearningCurves}),
with some slope $b$ typically more or less a constant for different unique representations and related to the effective dimensionality of the problem,
and some off-set $\log{a}$, which typically reflects the capability of the representation to capture the most relevant feature variations in kernel space.
More specifically, the off-set measures the degree to which the representation encodes the right physics. 
An illustrative example for this statement can be given by comparison of the learning curves obtained for the CM representation versus derived matrices with off-diagonal entries dependent on alternative interatomic power-laws~\cite{BAML}.
For interatomic off-diagonal elements approaching London's $R^{-6}$ law, the representation achieved lower off-sets than for 
off-diagonal elements decaying according to Coulomb's law. 
Correspondingly, representation matrices with off-diagonal elements linearly or quadratically 
growing with the interatomic distances resulted in LCs with dramatically increased off-sets~\cite{BAML}.
At first glance, it might seem to be a given that the slope of LC (aka, ``learning rate'' of QML model) barely changes, 
when switching from one unique representation to another. 
It is therefore natural to ask if it is impossible to further speed up the learning process  as indicated by the dotted-dashed learning curve in Fig.~\ref{fig:LearningCurves}, exhibiting a much steeper slope.
Through an expert-informed reduction of effective dimensionality (through a priori removal of irrelevant information stored in randomly selected training data)
it was shown that this is indeed possible. 
Such strategies for a more rational sampling of training data will be discussed below in Section~\ref{sec:ALML}.}
Note that in contrast to conventional curve fitting, training errors for properly trained machine learning models applied to synthetic data are typically orders of magnitude smaller than the variance of the signal. As such, they are negligible and carry little meaning since noise levels are typically close to zero or at least many orders of magnitude smaller than label variance.
Consequences of model construction, i.e.~choice of regressor, 
metric, optimizer, loss-function, representation, or computational efficiency, 
become immediately apparent in the characteristic shape of learning curves. 
When training a small parametric regressor, e.g.~a shallow neural network
with few neurons, to estimate a complex and high-dimensional target function, 
the learning curve will rapidly `saturate' and converge towards a finite
optimal residual prediction error that can no longer be lowered by mere
increase of training set size. 
As such, it should come as no surprise that learning curves have emerged as a crucial tool for development, validation, comparison,
and demonstration purposes of QML models in the field.

\subsection{Loss-functions}
Imposing differential relationships during training amounts to adaptation of the loss-function to better reflect the problem at hand. In particular, inclusion of derivative information (gradients and Hessian)
has led to dramatic improvements when tackling the 
problem of potential energy surface fitting~\cite{CovariantKernelsSandro2016,chmiela2017machine,chmiela2018towards,chmiela2019sgdml}. 
A generalization of this idea to adapt the loss function for response properties of any QM observable was established for KRR in 2018~\cite{OQML,OQMLchimia,FCHL19}
(exemplified for forces, Hessians, dipolemoments, and IR spectra) and for deep neural nets in 2020 (FieldSchNet exemplified moreover also for solvent effects and magnetic effects)~\cite{gastegger2020FieldSchNet}.

While conventional machine learning assumes that train and test loss-function are identical, for atomistic simulation (or other application domains for that matter) a mathematically, more `gready', alternative might exist.
In particular, the role of gradients in loss-functions differing for training and testing has been studied in Ref.~\cite{christensen2020gradientsloss} with results suggesting that for predicting atomization energies throughout a CCS of distorted structures inclusion of gradients in training improves learning curves negligibly while surely inflating the number of necessary kernel basis-functions. However, when it comes to predicting the potential energy surface of a given system, they do improve the energy predictions in the above referenced studies.   
Conversely, when predicting gradients throughout CCS, 
the use of energies alone in training offers no advantage over using forces, suggesting that the inclusion of forces (if computationally less demanding than energies) should always be beneficial.  

\section{Representations}
One could consider the choice of functional form of the representation $\bf M$ to be part of the machine learning methodology. 
However, this is a much studied question which is at the heart of how one views CCS. 
More specifically, what are the truly defining aspects in a compound? And how does one measure similarity? These are old questions which have already been answered for an impressive 
array of applications and instruct much of the basic and fundamental text-book knowledge. 
For example, Hammett's $\sigma$-parameter provides a low-dimensional quantitative data-driven measure of similarity between distinct functional groups in terms of their impact on reaction rates or yields~\cite{Hammett_jacs1937,bragato2020Hammett}.
Within QML, physically more motivated representations are sought after for subsequent use within high-dimensional non-linear interpolators which are more universal and transferable. 
As illustrated for KRR in Fig.~\ref{fig:kernel}, also the specific form of the representation (as well as the metric used) can dramatically affect the way CCS is represented within the regressor.
It should therefore not come as a surprise that the data-efficiency of QML models was found to depend dramatically on the specifices of the representation used. 
Since the importance of the choice of the distance measures has already been mentioned above, in this section we will focus on research that was done to find improved representations.

The choice of this particular compound representation, 
a.k.a.~descriptor or feature, plays a particularly crucial role. 
Correspondingly, it should come at no surprise that substantial research on the design
of descriptors has already been made in the fields of chem-, bio-, or materials 
informatics where scarce data is typical~\cite{DescriptroOverviewMeringer2005,TodeschiniConsonniHandbookDescriptor}.
Often, a large set of prospective features is hypothesized, and subsequently
reduced within iterative procedures in order to distill the most relevant variables
and low-dimensional projections pertinent to the problem at hand (see above). 
While it is certainly possible to also pursue this approach within a quantum mechanical
description of CCS~\cite{ghiringhelli2015big}, its heuristic and speculative character remains 
as unsatisfactory as its lack of universality and transferability.
Fortunately, the quantum nature of CCS allows us to follow more systematic and rigorous procedures
in order to address this question.

For example, it is a necessary condition for any successful ML model
to rely on uniqueness (or completeness)  in the representation, as pointed out, proven and discussed several years ago in Refs.~\cite{MoussaComment,FourierDescriptor},
and more recently in Refs.~\cite{ceriotti2020completeness,parsaeifard2020assessment}.
Uniqueness is essential in order to avoid the introduction of spurious noise
due to uncontrolled `coarsening' of that subset of degrees of freedom which is neglected.
Molecular graphs based on covalent bond connectivity only, 
for example, do not account for conformational degrees of freedom.
Consequently, their use as a representation will make it impossible to
quench prediction errors below the variance of the 
target property's conformational distribution
---no matter how large the training set.

Other characteristics, desirable for representations to display, include compactness, computational efficiency, symmetries, invariances, and meaning. 
Representations, in conjunction with the regressor's functional form, 
define the basis functions in which properties are being expanded, 
and strongly affect the shape of the learning curves.
E.g.~accounting for a target property's invariances through the representation
typically leads to an immediate decrease of the learning curve's off-set.

While it is possible to model all QM properties using the same representation and kernel~\cite{SingleKernel2015},
as also demonstrated for neural nets with multiple outputs already in 2013~\cite{Montavon2013}. 
It should be stressed, however, that this is a distinct feature of QML which stands in 
stark contrast to conventional QSAR or QSPR where the ML model is typically
strongly dependent on the target property.
If regressor, metric and representation ${\bf M}$ are independent 
of the label, i.e.~the quantum property, there is a strict analogy to quantum mechanics
in the sense of the Hamiltonian (or the wavefunction) of a system not 
depending on the operator for which the expectation value of any given observable is 
calculated~\cite{QMLessayAnatole}. 
This becomes obvious by considering the training of a KRR model 
where the regression coefficients are obtained through inversion of 
the kernel matrix, 
\bea
{\bm \alpha} & = & ({\bf K}+\lambda {\bf I})^{-1} {\bf p}^{\rm ref},
\eea
where for synthetic calculated data with signals being orders of
magnitude smaller than noise, the regularization $\lambda$ 
(also known as noise-level) is typically close to zero. 
Using property independent representations, metrics, and kernel functions 
it is therefore obvious that the regression coefficients adapt to each property only
because of the reference property vector ${\bf p}^{\rm ref}$. 
In Ref.~\cite{SingleKernel2015}, this has been illustrated numerically by
generating learning curves for various properties using always the same inverted
kernel matrix for any given training set size.

The predictive accuracy for specific properties varies wildly as a function of 
representation and regressor choice~\cite{googlePaper2017}.
The historic development over years 2012 to 2018 for a selection of ever 
improved machine learning models (due to improved representations and/or regressor architectures) 
can be exemplified for the prediction errors of atomization energies 
stored in the QM9 data set~\cite{QM9}
and has also recently been summarized in the context of the ``QM9-IPAM-challenge''
in Refs.~\cite{faber2020lecturenotesphysics,Anatole2020NatureReview,Burke2020retrospective}. 

The inclusion of increasingly more (less) `physics' in the representation 
has been demonstrated to systematically improve (worsen) learning curves~\cite{BAML},
and has been followed by a series of developments which have all been benchmarked on the same set of
atomization energies of small organic molecules in the QM9 dataset~\cite{QM9}, and which demonstrate the progress made. While binding energies of `frozen' geometries still constitute an application somewhat remote from most real-world applications in chemistry, from a basic physics point of view they do represent a crucial intermediate step before tackling more complex properties. In other words, if machine learning models failed to predict binding energies, one should not expect them to work for free energies. 
But also from a practical point of view, the computational cost of single-point energy calculations typically dominates all quantum chemistry compute campaigns, and therefore represents one of the most worthwhile targets for surrogate models used for the navigation of CCS.

We also note that with the emergence of deep neural networks, the problem
of also 'learning' the representation can be mitigated to be incorporated
in the overall learning problem~\cite{NeuralMessagePassing,SchNet}.
While many intriguing  and sophisticated representations, such as 
Fourier-series expansions~\cite{FourierDescriptor},
wavelets~\cite{eickenberg2018solid}
multi-tensors~\cite{Shapeev_AL_2018}, 
or
molecular orbitals~\cite{welborn2018transferability}
have been proposed,
most representations can be categorized to either 
correspond to adjacency matrices or to many-body expansions through distribution functions. 
We therefore limit ourselves to predominantly discuss in the following both 
predominantly in the context of KRR based QML models. 
A comprehensive overview on representation for KRR based QML models has also recently been contributed by Rupp and co-workers~\cite{Rupp2020Review}. 

\subsection{\textcolor{blue}{Discrete}} \label{sec:x_descrete} 
Coordinate-free, bonding neighbors (covalently bonded atom pairs) based graphs, 
as well as their systematic extensions to arbitrary number of neighboring shells, 
have formed an important research direction in cheminformatics for many 
years~\cite{SignatureFaulon2003,ecfp,CanonizerMeringer2004,DescriptroOverviewMeringer2005,TodeschiniConsonniHandbookDescriptor}. 
In 2011, supervised learning was proposed as an alternative to 
solving Schr\"odinger's equation throughout 
a chemical compound space relying as a representation on a complete undirected labeled graph 
that encodes the simplex spanned by all atoms~\cite{CMarxiv,CM}. 
More specifically, this graph was represented by the `Coulomb matrix' (CM),
an atom by atom  matrix with the
nuclear Coulomb repulsion on off-diagonal elements, 
and with approximate energy estimates of free atoms
($E_I \approx 0.5 Z_I^{2.4}$~\cite{englert1988semiclassical}) as diagonal elements.
Formal requirements such as uniqueness, translational and rotational invariance, as well
as basic symmetry relations (symmetric atoms will share the same matrix elements
in their respective rows or columns) are all met by the CM.
Atom index invariance can be achieved through use of 
its eigenvalues (thereby sacrificing uniqueness~\cite{MoussaComment,MoussaReply}), 
sets of randomly permuted CMs~\cite{Montavon2013},
or sorting by norms of rows~\cite{AssessmentMLJCTC2013}
-thereby loosing differentiability due to sudden switches in ranks~\cite{onur2020wasserstein}).
We reiterate once more that the atom indexing dependence 
can be mitigated through use more sophisticated distance 
measures such as the Wasserstein metric~\cite{onur2020wasserstein}.

Similarly encouraging findings of KRR based QML models applicable throughout
CCS were quickly reproduced for other materials classes 
such as polymers~\cite{ML4Polymers_Rampi2013}, 
or crystalline solids~\cite{GrossMLCrystals2014}.
While off-diagonal elements with a London dispersion power law ($r^{-6}$) 
have subsequently been found to be preferable for QML models of atomization energies~\cite{BAML},
other representations ({\em vide infra}) offer lower learning curve off-sets.
In particular, the bag of bonds representation (BoB) is worthwhile mentioning~\cite{BoB}. 
Introduced in 2015, BoB groups the entries of the CM in 
separate sets for each combination of atomic element pairs in bonds within
which the elements have been sorted. 
When calculating the similarity between two molecules, 
only Coulomb repulsions between atoms with the same nuclear charge are being compared, 
rendering thereby the similarity measurement more balanced, and effectively lowering the learning curve off-set. 
While even more compact than the CM, BoB lacks uniqueness due to being strictly
a 2-body representation which can not distinguish between homo-metric configurations~\cite{FourierDescriptor}. 
The generalization of BoB towards the explicit incorporation of covalent bond information, angles, as well as dihedrals in terms of a systematic expansion in Bond, Angle, and higher order interactions (i.e., BAML representation) was accomplished in 2016~\cite{BAML} using functional forms and parameters from the universal force-field~\cite{UFFRappeGoddard1992}.
\textcolor{blue}{A similar, but more elaborate, parameter-free, many-body dispersion (MBD) based 
representation involving two and three body terms~\cite{pronobis2018many} was proposed later in 2018.}

The CM \textcolor{red}{has been} proved essential as a base-line for the interpretation, analysis, and further development
of subsequent QML models. 
It has also been adapted successfully to account for periodicity in the condensed phase, 
as evinced by learning curves for formation energies of solids~\cite{MLcrystals_Felix2015}.
For other properties, such as forces, electronic eigenvalues, or excited states, 
the CM (or its inverse distance analogues for QML applications with fixed chemical composition) 
is still competitive with state of the art 
representations~\cite{ML-TDDFTEnrico2015,HDAD,chmiela2017machine,
chmiela2018towards,chmiela2019sgdml,stuke2019chemical,ghosh2019deep,westermayr2019machine,
sauceda2020GDML}.
Furthermore, due to its uniqueness, compactness, and obvious meaning, 
the CM (or its variants) are conveniently used to overcome frequent 
data analysis problems in atomistic simulations, such as
removal of duplicates, quantification of noise-levels, simple learning tests.

Regarding the interatomic distance dependent decay rate of off-diagonal elements, 
it is also worthwhile to mention exponential functions, rather than $1/r$. 
In particular, the overlap matrix between atomic basis-functions of 
all atoms has been proposed~\cite{OverlapMatrix2016},
and used with great success for QML models of 
basis-set effects~\cite{schuett2018MLbasisCP2K}
and excited states surfaces~\cite{OM4excited2020}.
The overlap matrix was also included within a recent sensitivity assessment of various state-of-the art representations and performed in impressive 
ways~\cite{parsaeifard2020assessment}.
A constant-size descriptor based on a combination of the CM with more 
common molecular graph fingerprints 
was also proposed in 2018~\cite{collins2018constant}.

Viewing BoB and CM as first and second rank tensors, 
to the best of our knowledge, 
use of a third rank tensor 
(explicitly encoding the surface of all possible triangles in a compound)
has not yet been tested.

\subsection{\textcolor{blue}{Continuous}} 
\textcolor{red}{Aforementioned discrete and global representations such as BoB enjoy fast computation.
One important requirement for this kind of representation to work, however, is to introduce atom indexing invariance by 
sorting atoms according to the magnitude of entries belonging to each bond or other many-body types.
This is artificial and may introduce derivative discontinuities with unfavorable consequences in related applications
such as force predictions.}

\textcolor{red}{The sorting and associated problems can be naturally overcome by selecting continuous or distribution based representations, which, in essence, 
integrate out atom index dependent terms such as
 distance (w/wo angle and dihedral angle) and/or nuclear charge (i.e., alchemically~\cite{FCHL})
through use of smeared out projections (a Gaussian is usually put on each degree of freedom).}
Distribution based representations, also closely related to many-body 
or cluster expansion~\cite{drautz2019atomic},
have gained much popularity also within QML models building on Behler's seminal work on
atom-centered symmetry functions for training neural networks on potential
energy surfaces~\cite{Neuralnetworks_Behler2011}, 
or through the subsequent introduction of smooth overlap of atomic potentials (SOAP) 
for use in GPR by Bartok et al. in 2013~\cite{BartokGabor_Descriptors2013}.
The first variant of linearly independent distribution based representations for QML models, applicable
throughout CCS, a Fourier series expansion of nuclear charge weighted
radial distribution functions, was also contributed already in 2013~\cite{FourierDescriptor}, 
albeit published in its final version only in 2015.
Radial distributions were also used for representing crystals in solids
in 2014~\cite{GrossMLCrystals2014}.
The atomic spectrum of London Axilrod-Teller-Muto (aSLATM) terms was first
presented in 2017 within the `AMON' approach by Huang et al.~\cite{Amons_nchem2020} ({\em vide infra}), 
yielding unprecedentedly low off-sets
in learning curves for atomization energies in the QM9 data-set~\cite{QM9}. 
In that same year, SOAP based QML models were generalized and shown
to be also applicable throughout CCS~\cite{CeriottiScienceUnified2017}. 

\begin{figure}
\includegraphics[scale=0.5]{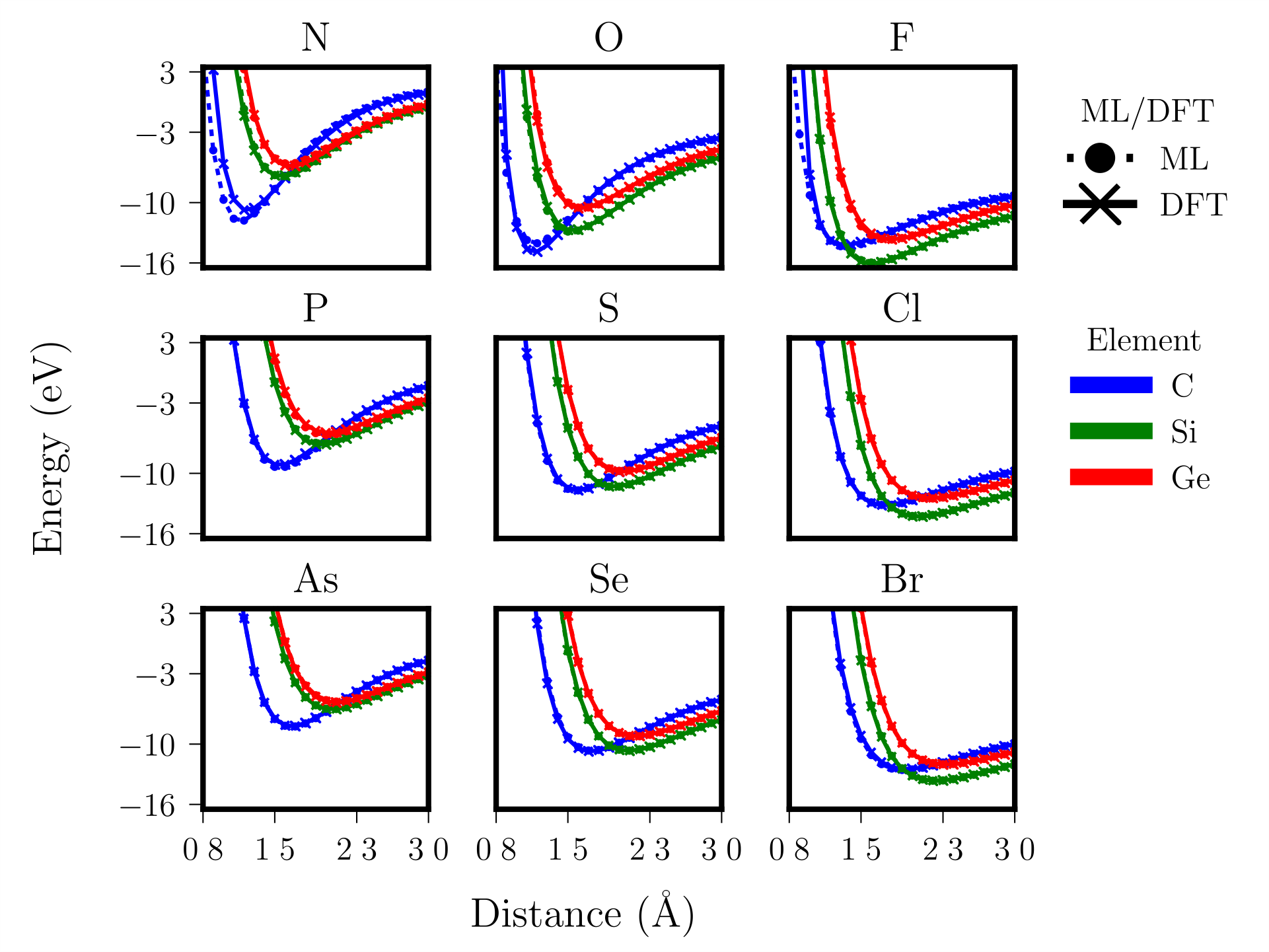}
\caption{\label{fig:FCHL}
QML models infer properties for new chemical compositions.
DFT and QML (FCHL+KRR) based predictions of covalent triple, double, and single bonding 
between group IV and V (left column), VI (mid column), and VII (right column) elements, respectively.
Open valencies in the group IV elements have been saturated with hydrogens.
QML models were trained on the DFT results for all those chemical elements 
that are not present in the query molecule. 
Reproduced from 
Faber, Christensen, Huang, von Lilienfeld, {\em J. Chem. Phys.}, {\bf 148}, 241717 (2018)~\cite{FCHL}; 
licensed under a Creative Commons Attribution (CC BY) license.
}
\end{figure}

The generic histogram of distances, angles, and dihedrals (HDAD)~\cite{HDAD}, \textcolor{red}{could be regarded as a continuous but simplified version of BAML,
both including many-body terms up to torsions,}
was contributed in 2017. 
In the following year, Faber et al.~conceived \textcolor{blue}{the idea of adding
alchemical degrees of freedom in} a structural distribution based many-body representation,
dubbed FCHL18~\cite{FCHL} (FCHL indicating the first letters of the last names of the authors and 18 the year 2018).
The FCHL family of representations encodes a systematic interatomic
many-body expansion in terms of Gaussians weighted by power-laws,
due to the insights gained in Ref.~\cite{BAML}. 
Power law exponents and Gaussian widths were optimized as hyper-parameters 
through nested cross-validation during training.
\textcolor{blue}{FCHL18 consists of three parts:} 
The \textcolor{blue}{one-body} term corresponds to a 2-dimensional Gaussian encoding
the chemical identity of the atom in terms of groups and periods of the 
periodic table;
The \textcolor{blue}{two-body} term encodes the interatomic distance distribution scaled
down by $r^{-4}$, and the \textcolor{blue}{three-body} term encodes all angular distributions 
and is scaled down by $r^{-2}$. 
The impact of four-body terms has been tested but was found to have negligible impact on learning curves~\cite{FCHL}.
Most importantly within the context of CCS, FCHL18 based QML models have 
been demonstrated to be capable of accurately inferring property estimates 
of systems containing chemical elements which were {\em not} part of training.
\textcolor{red}{More specifically, consider the family of molecules of formula H$_n$Y$\sim$X, 
where Y corresponds to an element from group IV (either C, Si, or Ge); 
where `$\sim$' represents single, double, or triple bond depending on chemical element X being from group VII, VI, or V, respectively;
and where $n$ is the number of H atoms that saturates the total valences.
Semi-quantitative covalent bond potential binding curves have been predicted for any X/Y/bond-order combination using QML models
after training on corresponding DFT curves for all other molecules that neither contain X nor Y.}
(see the top- and left-most subplot in Fig.~\ref{fig:FCHL} for an illustration).
\textcolor{red}{For example, the ML binding curve of HC\#N was obtained after training on binding curves 
of all other molecules that neither contained N, nor C, i.e.~when predicting the blue curve in the upper left
panel of Fig.~\ref{fig:FCHL}, the red and green curves of that panel were not part of training, nor any other blue
curve from the other panels. }
FCHL19, a recent revision has been shown to provide a substantial speed-up in training and testing, 
while imposing only a small reduction in predictive accuracy~\cite{FCHL19}.

We note in passing also the related moment tensor model (MTM) by Shapeev and co-workers, introduced in 2018~\cite{gubaev2018machine}, 
as well as the unifying interpretation of many of the popular distribution based representations
by Ceriotti and co-workers~\cite{WillatCeriotti2019unifyingrepresentations}.



\section{Regressor}
Depending on how regression parameters are being obtained, 
the incorporation of legacy methods in QML models applicable throughout CCS is 
typically done either within neural networks or within Gaussian process regression (GPR) (or kernel ridge regression, KRR for short). 
Here, we mostly focus on kernel methods, 
mentioning only shortly 
the idea of transfer learning in neural network models~\cite{taylor2009TransferLearning},
which is also applicable to QML models as shown in 2018 by
Smith et al.~\cite{OutsmartingQuantumChemistry}.

More specifically, five categories of QML models can easily be distinguished,
each of which accounting for legacy information in its own way: 
QML models of parameters of existing models, QML models of corrections to
existing models ($\Delta$-ML), 
multi-fidelity ML (MF-ML), 
multi-level-grid-combination (MLGC), and
transfer learning techniques.
We briefly review each of these in the following.

\subsection{ML models of parameters}
Existing force-field models can capture nicely the essential physics of a wide range of chemical systems, the main drawback being that force-field parameters (e.g., atom charges, harmonic force constants, etc.) are often rigid and unable to adapt to different atomic environments. Therefore, it would be natural to make these parameters flexible and predicted by ML models. 
This idea dates back to the nineties and the first piece of related works was done by Hobday et al.~\cite{hobday1999applications}, where they proposed a neural network model to predict parameters of the Tersoff potential for C-H systems. In 2009, Handley and Popelier proposed to use machine learning models for 
multipole moments~\cite{MachineLearningWaterPotential_Handley2009}.
This idea was revisited in 2015, when learning curves for atomic QML models 
of electrostatic properties, such as atomic charges, dipole-moments, 
or atomic polarizabilities were presented~\cite{MTP_Tristan2015}. 
Their use for the construction of universal
non-covalent potentials was established in 2018~\cite{bereau2018non}. 
Neural-network based equilibrated atomic charges were also proposed in 2015 
by Goedecker and co-workers~\cite{Goedecker2015NN,Goedecker2017neuralnet}, 
and in 2018 by Roitberg, Tretiak, 
Isayev and co-workers~\cite{Isayev2018charges,Nebgen2018charges}.

Similar strategy could also be applied to semi-empirical quantum chemistry methods relying on parameters typically fitted by computational/experimental data. In 2015, QML models of nuclear screening parameters were contributed by Pavlo and co-workers~\cite{Pavlo2015parameterML}. 
In 2018, unsupervised learning for improved repulsion in tight-binding DFT 
was introduced by Elstner et al.~\cite{elstner2018unsupervisedTightBinding},
followed by substantial further improvements in 2020~\cite{Tkatchenko2020repulsionTightBindingDFT}. 
Extended H\"uckel theory was revisited in 2019 by 
Tretiak and co-workers~\cite{zubatyuk2019machine}.

\subsection{$\Delta$-ML}
The idea to present QML models of label corrections applicable throughout CCS and 
which systematically improve with training data size
was first established in 2015 in terms of $\Delta$-machine learning.
Numerical results provided overwhelming evidence for the success of this idea
as demonstrated for modeling energy and geometry differences between various 
levels of theory including PM7, PBE, BLYP, B3LYP, PBE0, G2MP4, HF, MP2, CCSD, and CCSD(T)
for QM9~\cite{QM9} and subsets thereof~\cite{DeltaPaper2015}.

$\Delta$-ML also works for correcting complex and subtle properties, such as
van der Waals interactions in extremely data-scarce limits as illustrated
for DFT corrections based on training sets with less than one hundred
training instances~\cite{mezei2020noncovalent}, 
or to model higher order corrections to alchemical perturbation density 
functional theory based estimates of heterogeneous catalyst 
activity~\cite{griego2020machinealchemy}.
Among many other applications, $\Delta$-ML has enabled corrections to 
electron densities~\cite{Kieron2020MLDFT}, 
electron correlation based on electronic structure representations within Hartree-Fock~ or 
MP2 level of theory~\cite{Vogiatzis2020transferableMP2basedMLofCC}, 
or DFT and CCSD(T) based potential energy surface estimates~\cite{Bowman2020delta}.

For noise-free data and functional QML models (unique representations),
numerical results for learning curves indicate a constant lowering of 
off-set, no matter which training set size. 
Such non-vanishing improvement appears to turn into vanishing improvement
when employing $\Delta$-ML in order to correct 
low-quality or coarse-grained baselines, 
such as a semi-empirical PM7~\cite{DeltaPaper2015},
or Hammett's relation~\cite{bragato2020Hammett}.

\subsection{Multi-fidelity} 
The success of $\Delta$-ML is encouraging, enabling possible a significant reduction in high-accuracy reference quantum chemical data necessary for training, to reach the same level of predictive accuracy as traditional QML models.
However, it still consumes a considerable amount of data calculated at some high level of theory, as its structure in design is too simple to fully exploit the underlying correlation between varied quality of properties.
In fact, well-established quantum chemical methods abound in literature, exploiting effectively the underlying correlation, in the name of the so-called composite methods, for example, the famous G$n$ series~\cite{PopleG2,PopleG3,PopleG4}.
In essence, these methods approximate some specific part of correlation energy (e.g., energy lowering due to inclusion of diffuse orbital in basis set) from a high level of theory (for instance CCSD(T)) by the same quantity calculated from a relatively low level of theory (say MP2).
Due to error cancellation, composite methods have been proven to be extremely effective towards reaching an accuracy of experimental quality and are widely used for calculations of high-quality thermochemical data~\cite{PopleG2,PopleG3,PopleG4}.

To do interpolation and meanwhile exploit error cancellation effectively, 
multi-fidelity ML (MF-ML) comes into play.
The core idea of MF-ML is hereafter demonstrated by total energy ($E$) prediction. For brevity, we deal with two levels of theory (the low and high level are denoted by 0 and 1 respectively) and focus on one flavor of MF-ML, i.e., recursive KRR (r-KRR for short, or MF-KRR)~\cite{zaspel2018boosting}, which is similar to its counterpart, recursive GPR (r-GPR, or MF-GPR)~\cite{gratiet_recursive_gp_2014,kennedy_autoregressive_gp_2000} and differs to MF-GPR to some extent, in analogy to the difference between KRR and GPR.
Unlike $\Delta$-ML, MF-ML comprises multiple machines with different labels to learn (two for our exemplified case).
The first one is just a traditional QML model trained on a set of data (denoted as $S_0$) associated with the low level of theory, i.e., $E_j^0=\sum_{i\in S_0} c_i^0 k(\fatx_i, \fatx_j)$, where $j\in S_0$, $\fatx$ denotes the molecular representation vector and $c^0$ is the regression coefficient associated to the low level of theory. This machine is also called the baseline model.
Then we build a second machine with training set $S_1$ satisfying $S_1\subset S_0$ and energy delta, i.e., $E^1 - E^0$, as label, the same as a $\Delta$-ML model. In math, $\Delta E_n^{0\rightarrow 1} = E_n^1 - E_n^0=\sum_{m\in S_1} c_m^1 k(\fatx_m, \fatx_n)$, where $n\in S_1$. Once trained separately for each machine, all $c^l$'s are obtained and MF-KRR predicts the property of any query $q$ out-of-sample at the high level of theory by $E_q^1 = E_q^0 + \Delta E_q^{0\rightarrow 1}$.

Extending r-KRR to more than two levels of theory is straightforward: except the baseline model for the lowest level, one needs to build one machine for every two adjacent levels of theory 
and the final test energy is just the summation of the inferred energies by all machines, i.e.,  $E^L_q = E^0_q + \sum_{l=0}^{L-1} \Delta E^{l\rightarrow l+1}_q$, where $l$ is the level indicator (starts from 0, the lowest level) and $L$ corresponds to the largest $l$, or the target level. Bear in mind that $S_0\subset S_1\subset \dots \subset S_L$.

We note by passing that MF-GPR has a rather different formulation compared to MF-KRR and benefits from the stochastic nature of GP, i.e., it is capable of providing the variance estimate of prediction.
Like GPR, data at each level of theory in MF-GPR is modeled as a GP~\cite{kennedy_autoregressive_gp_2000,gratiet_recursive_gp_2014}, and every two adjacent levels is connected by a linear transformation, i.e., $E^{l+1} = \gamma E^{l} + \epsilon$, where $\gamma$ is a scaling factor and $\epsilon$ is a correction term respectively and both of which may depend on the two involved levels of theory (i.e., $l$ and $l+1$).
Nevertheless, both MF-KRR and MF-GPR could end up with the same set of working equations under certain conditions. For detailed derivation of the equations of MF-GPR, the reader is referred to an early review on QML~\cite{huang2018quantum}.
Last but not least, one should note that MF-KRR converges towards the conventional KRR model 
associated with the highest level as the difference between training sets for each machine vanishes.

Albeit well-founded in mathematics decades ago, 
the power of MF-ML has not been harvested until recently.
Applications include quantum collision for the 
Ar-C$_6$H$_6$ system by Cui et al.~\cite{cui_mfml_quantum_collision_2015}, 
bandgap prediction of solids done by Pilania et al.~\cite{pilania_mfml_bandgap_2017}, 
dopant formation energy prediction in hafnia by Batra et al.,~\cite{batra2019multifidelity},
high-accuracy potential energy surface prediction for 
small molecules by Wiens et al.~\cite{wiens_mfml_Esurf_2019},
and the recently performed molecular crystal structure prediction study 
by Egorova et al.~\cite{egorova_mfml_molcrystal_struct_pred_2020}

\subsection{Multi-level-grid-combination} 
In spite of the drastic improvement over $\Delta$-ML, MF-ML has its own limitations. 
For one, the computational cost of the base-line evaluation for every query compound
can still be considerable.
Furthermore, it must be strictly satisfied that the increasingly more expensive 
training sets form a nested structure, implying that possible and beneficial correlations 
between non-nested reference data calculated at different level of theory are not being exploited.
To overcome this drawback, Zaspel et al.~\cite{zaspel2018boosting} proposed a multi-level model in 2018, 
combining successfully ML with sparse grid (SG)~\cite{garcke_data_2001}, a numerical technique widely used to integrate/interpolate high dimensional functions.

The genuine SG approach assumes (quasi-)uniform grids along each dimension, which serves as basis functions (more precisely, centers of basis functions, such as triangular function) and based on tensor products of which any multidimensional function could be represented/expanded~\cite{garcke_data_2001}. The expansion weight for each tensor product, is dependent on only the indices of associated grid and spacing along each dimension, and determined by multi-variant Boolean algorithm~\cite{delvos_multivariate_boolean_1982}.

Replacing such grid by abstract variable (or combinations of which) such as electron correlation level ($x_C$), basis set ($x_B$) and expressing system property as a function of these abstract variables is definitely a refreshing strategy.
For example, the total energy of a system could be expressed as $E=E(x_C, x_B$).
Given some sparse grids comprising small $x_C$ combined with all $x_B$'s, and small $x_B$ combined with all $x_C$'s, and intermediate $x_B$'s combined with intermediate $x_C$'s, we are able to interpolate/extrapolate the $E$ at some different combination of $x_C$ and $x_B$.
Of particular interest is extrapolation to regions unsampled, i.e., regions with large $x_C$ and $x_B$.
However, one major issue with such extension is the elusive nature of distance between two abstract variables, which is essential in determining the weight associated with each grid, as mentioned above.
More specifically, it is unknown how to quantitatively characterize how distant HF and MP2 are along the dimension $C$,
though qualitatively it it certain that HF lies closer to MP2 compared to CCSD(T).
The $B$ subspace, is understood much better, as the magnitude of $x_B$ could be at least characterized by the largest angular channel, or more straightforwardly -- though less rigorously -- by the number of basis functions.
This ill definition of these abstract variables, is totally absent in genuine SG as grids there reside typically in Euclidean space and therefore distance is well-defined.
To rise to this problem, a workaround is to assume uniformality of grids along each dimension (i.e., equidistant) and grids along each dimension is represented simply by indices starting from 0 (now weights depend solely on the indices of grids).
This, however, should always be done with great care.
In the original MLGC paper~\cite{zaspel2018boosting}, electron correlation levels are reasonably chosen as HF, MP2 and CCSD(T), together with three basis sets, i.e., STO-3G, 6-31G and cc-pVDZ (the number of basis functions increases by a factor of $\sim$2).

Note that the aforementioned SG approach deals with typically one system at a time.
To incorporate it within ML framework, one extra variable has to be introduced, i.e., training set ($x_N$), the size of which indicates the magnitude of $x_N$~\cite{zaspel2018boosting}. Accordingly, $E=E(x_C, x_B, x_N)$.
Unlike subspace $C$ or $B$, $x_N$ is well-defined with explicit value.
Nevertheless, it has to be treated in a similar fashion as for $x_C$/$x_B$.
I.e., given the minimal $x_N$ (aka. $N_0$) and a ratio ($s$) between any two adjacent $x_N$'s, training sets are to be assigned an array of indices starting from 0 (for $N_0$) and an increment of 1.
This assignment is necessary so as to comply to what has been done for subspaces $C$ and $B$.
Now each grid in this abstract space is a combination of three variables: ($x_C, x_B, x_N$), with $x_I \in \{0, 1, \dots, I_{max}\}, I\in \{C,B,N\}$.
For each such combination, an associated ML model is trained (with $N$ training instances of course). Given a query system, its energy is predicted as a weighted summation of test energies from all ML models, with weights derived from Boolean algorithm~\cite{delvos_multivariate_boolean_1982,zaspel2018boosting}.
Note that in practise, to reduce the cost of generation of reference quantum data, a large $x_N$ is associated with a low level of correlation and/or small basis set, while only few(er) labeled data is needed for high(er) correlation level and/or large(r) basis set.

With the above setting, Zaspel et al. were able to show~\cite{zaspel2018boosting} that MLGC enables $\sim$10 fold of reduction (cf. traditional single level ML model) in the costly highest level quantum data (i.e., CCSD(T)/cc-pVDZ) to reach chemical accuracy in predicting atomization energy of out-of-sample QM7b molecules.
Last but not least it is worth pointing out that MLGC could be viewed as an extension 
of MF-ML if only one dimension is being considered. 

\subsection{Transfer learning}

While multi-level methods are most naturally combined with kernel methods, they may have even more far-reaching effects for neural network (NN) models,
as training a NN model, deep NN (DNN) in particular, is 
a non-trivial problem. 
\textcolor{blue}{Furthermore}, current ad-hoc DNN models are typically 
specialized, meaning a (D)NN model may need to be re-trained 
for a slightly different task.

Transfer learning (TL)~\cite{pan_survey_2010} is one popular approach employing multiple levels in machine learning that can greatly alleviate the aforementioned problems, which reuses the knowledge, gained through solving one task (base task), as a starting point for a second task (target task), different but highly related.
For instance, knowledge obtained from learning to infer DFT energies could be applied to infer CCSD(T) energies~\cite{smith_approaching_2019}.
A successful transfer of knowledge can improve the performance of the target DNN model significantly. Speaking the language of learning curve, TL could offer~\cite{pan_survey_2010}
i) smaller offset as the transferred model per se provides a decent starting point and
ii) steeper learning curve due to the transferred model usually represents a parameter space rather close to the optimal one.

Based on the type of traditional ML algorithms involved, TL could be categorized into several variants.
Here we focus on a variant named inductive transfer learning,
in which the labeled source and target domains are the same, yet the source and target tasks are different from each other.

In TL, there are two essential ingredients: i) a pre-trained model, obtained by either training a network from scratch on some data set and a specific task, or simply from published models; ii) target network, to be trained on a target data set and task, but utilising the learned features from i).
This process is likely to work only if the features are general (i.e., generic features) to both base and target tasks, 
as would be captured by the initial layers of NN models.
When re-training the target model, one may choose to either
freeze the initial layers in the base network to use them as feature extractors for the target model
or fine-tune the last several layers further for improved performance.
A rule of thumb is to freeze when target labels are scarce (to avoid overfitting), while fine-tune otherwise.

To develop a successful TL model, it is vital to choose the proper base model and associated training data set.
However, it remains largely an open question about how to make the choice and which may require profuse intuition developed via experience.
Furthermore, there exists one major potential risk of using TL, i.e., negative transfer,
which refers to scenarios where the reuse of base-task knowledge degrades the overall performance of the target task.
To avoid negative transfer, one may have to resort to approaches that explicitly model relationships between tasks
and include this information in the transfer method.~\cite{olivas_handbook_2009}

Applications of TL cover mainly computer sciences, such as image recognition and natural language processing. For chemistry-related problems, TL is emerging as a promising approach. Examples include Smith et al.'s work on predicting CCSD(T)/CBS energies based on transferred knowledge gained through training on DFT energies~\cite{smith_approaching_2019} for the ANI-1x data set (see section~\ref{sec:gdb}), Iovanac et al.'s work on property prediction of QM9 molecules~ \cite{Iovanac2020_TL_QM9_property}, as well as Cai et al.'s recent work on drug discovery~\cite{cai_transfer_2020}.

\section{Training set selection}\label{sec:ALML}
Among all factors determining the performance of a QML model, training set selection plays another fundamentally important role in the sense that all knowledge essential for making confident predictions are implicitly encoded in the training data.

\indent Several pertinent fundamental and distinct questions have remained open. 
\begin{itemize}
    \item[{\bf Q1}] how to extract the most representative and least redundant general subset from a given data set?
    \item[{\bf Q2}] how to quantitatively define the suitability of a given training set for a specific query at hand?
    \item[{\bf Q3}] how to systematically select the most relevant training set for a specific system?
\end{itemize}
Due to their highly nonlinear impact of training instances on model parameters, these questions are challenging, and have not been studied much.
Of course, the problem of training set selection is not a problem unique to chemistry, and it is relevant to most supervised learning problems in other fields. 
Currently, the aforementioned issues are mostly addressed through a common strategy called random selection. Though universally applicable, random selection suffer inevitably from selection bias inherent in the data itself.
More specifically, in the randomly selected training set, many instances can be ignored and their inclusion in training
does not improve predictive performance of the QML model (due to redundancy), or could even degrade it (due to being irrelevant for a given query test or due to noise).

Bias could become a very serious issue as the systems under study are increasingly more complicated.
The origins of the bias issue could be divided into two components: i) curse of dimensionality. This is mainly related to the size of the systems, and plagued further by compositional diversity. More specifically, as the system size and/or the encompassing number of types of elements grow, the size of the thus-spanned CCS grows combinatorially (see above). 
ii) the inhomogeneity of CCS. The energetics in chemistry typically favors one kind of bonding over another. For instance, hydrogen atom favors a single sigma bond with other atoms, while carbon atom can exhibit several different bonding patterns such as $sp^3$, $sp^2$ and $sp$. Consequently, random sampling will introduce more subsampling of hydrogen environments, but proportionally fewer C$-sp$ local environments leading to worse model estimates of properties for C than for H.

To tackle the bias issue, previous and ongoing research has been trying to almost exclusively tackle {\bf Q1},
assuming a pre-existing dataset (or a dataset that is straightforward to generate, e.g., in molecular dynamics).
Examples include the use of genetic algorithms (requiring labeled data to gradually expand the optimal training set)~\cite{Nick2016GA,jacobsen2018fly},
or ``active learning'' approaches~\cite{Shapeev_AL,Shapeev_AL_2018}, which selects the most representative subset ``on-the-fly'' from a given set of unlabeled configurations,
i.e., no quantum chemical data is needed for making decisions about whether or not a query configuration is redundant.
The AMONs concept proposed by the authors~\cite{Amons_nchem2020} partially resolves question {\bf Q3} (cf. {\bf Q1} \& {\bf Q2}), at the same time allowing for significant dimension reduction of CCS as well as the effective removal of statistical redundancy of training sets (see below for details). Other related work shifts the attention to training set reduction instead, primarily in molecular dynamics simulations, for instance, Li et al.~\cite{Zhenwei2015} proposed a ``learning-and-remembering'' scheme, in which the decision to recompute QM data for a new configuration was taken every $n$ steps. Another relevant contribution to active learning in CCS was made in 2018 by Smith et al.~\cite{smith2018less} relying on `query by committee', i.e.~ensemble information obtained through use of multiple neural networks (of the ANI kind~\cite{ANI_IsayevRoitberg2017}). 
Potentially promising alternative directions could possibly be inspired by recent developments in computer science, among many others notably 
the idea of artificial ``soft'' labels, curated through carefully blending features of training instances~\cite{wang2020_softlabel_2,sucholutsky2020_softlabel_1}.
In the original paper, this idea was tested on the MNIST data set (a database of handwritten digits) and similar performance was achieved with much fewer but soft labels, as compared to training on almost the entire data set. This idea should in principle also be applicable to CCS requiring the design of some fictitious averaged training molecules, interestingly probable to violate common rules of chemical bonding.
In the following, we review the three most promising approaches towards training set selection: genetic algorithm, active learning and the AMONs approach.

 \subsection{Genetic algorithm}
Genetic algorithms (GA) have been widely used in (global) optimization problems in quantum chemistry, such as first principles based global structure optimization~\cite{cerqueira2015materials} (for compounds with desired physio-chemical property), a key topic in the inverse-design problem~\cite{sanchez2018inverse}.
To the best of our knowledge, the first piece of work about using GA for training set selection within QML was done by reding et al.~\cite{Nick2016GA} for molecules, followed by Jacobsen's work~\cite{jacobsen2018fly} on SnO$_2$(110) surface reconstruction.

In the following, we discuss the central idea of GA for the selection of the most representative set of QM9 molecules as done in Ref~\cite{Nick2016GA}. For applications to other properties and systems such as chemisorption systems~\cite{GA4chemsorb}, only technical details will differ.
Given a set ($S_0$) of $N$ molecules, GA carries out three consecutive steps for optimization:
(a) Generate $M$ random sets of size $N_1$.
This forms a starting population of training sets (aka. the parent population), labeled as $\hat{s}^{(1)} = \{s_i\}$, where $i\in \{1,2,\dots,M\}$.
Note that the initial size needs to be balanced against the diversity of the molecules for optimal performance.
(b) Train a QML model on each set in $\hat{s}^{(1)}$,
and then test on some joint pre-selected out-of-sample molecules (i.e., not part of $\hat{s}^{(1)}$),
the resulting test error $\epsilon_i$ (measured by for instance mean absolute error) serves as a ``fitness'' indicator, characterizing how fit $s_i$ is as a training set (smaller $\epsilon_i$ means better fitness).
(c) Evolution of $\hat{s}^{(1)}$ takes place through three consecutive steps: selection, crossover and mutation.
In the selection step, decisions have to be made on which $s_i \in \hat{s}^{(1)}$ should be kept in the population to produce a temporary refined smaller set $\hat{t}^{(1)}$ and a set with larger fitness value means higher probability to be kept in $\hat{t}^{(1)}$).
Crossover step involves the update on $\hat{s}^{(1)}$ from $\hat{t}^{(1)}$ and the resulting new population is re-labeled as $\hat{s}^{(2)}$,
each of which is obtained by mixing molecules from two subsets of $\hat{t}^{(1)}$).
The last step mutation randomly modifies molecules in some subset of $\hat{s}^{(2)}$ to promote diversity, e.g., replace -NH$_2$ group by -CH$_3$.
To avoid introduction of chemical environments alien to the whole data set, the replacements in mutation have to be constrained locally in $S_0$.
(d) Go back to step (b) and repeat b-d  until there is no improvement in the population and the fitness value has no significant improvements for over $n$ iterations.
The final converged set corresponds to a ``optimal'' training set, and is labeled as $\hat{s}$. 

It is not a surprise that selected $\hat{s}$ should be able to represent all the typical atomic environments in $S_0$, and therefore a QML model trained on $\hat{s}$ warrants significantly improved test results in comparison to randomly drawn training sets. 
As the fitness value decreases during the GA iterations, the QML models ``tried'' out the sensitivity with respect to inclusion of certain training instances, and this can serve their systematic inclusion/exclusion.
The usefulness of the optimized set $\hat{s}$ has to be assessed by the generalizability of the QML model trained on $\hat{s}$ to new molecules absent in $S_0$.
Indeed, improved generalizability is observed for PubChem molecules compared to random sampling, as was reported in~\cite{Nick2016GA}.

In spite of its power for solving hard optimization problems, such as finding the optimal training set composition, the drawback of most GA implementations is also obvious: It typically relies on the availability of labeled data to evaluate the fitness in each iteration.
As such, it only offers computational cost savings in terms of QML model efficiency, and not in terms of reducing the total need for available training data. 
Possible solution to circumvent this is to introduce heuristics in feature space, e.g.~accounting for the fitness by some distance metric instead, meanwhile avoiding the costly training-test procedure in each iteration~\cite{Kulik_TMC_NN_2020}.

 \subsection{Active learning}
Active learning (AL) is more interesting than GA for training set selection, as it can use directly unlabeled data, i.e.~{\em before} the acquisition of costly labels.
Intuitively, it makes sense that this should be possible as the quantum properties of any compound are implicit functions of its composition and geometry which is the only input required for calculating rigorous representations.
Among the many categories of AL algorithms used for determining which unlabeled data points should be labeled, below we mainly focus on the variance reduction query strategy, which labels only those points that would minimize output variance (uncertainty in prediction).
Note that the task of variance estimation is fundamentally different to mean error estimation, variance based selection method differ significantly from mean error based selection method (such as GA mentioned above) accordingly.
Relevant works on active learning include the $D$-optimality approach~\cite{Shapeev_AL,Shapeev_AL_2018} and methods based on variance estimators using Gaussian process regression (GPR)~\cite{ML4Kieron2012,Reiher2018_GPvar,Reiher2019_GPvar_Disp}, as well as neural network (NN) models.

Rooted in linear algebra, the \textit{D}-optimality approach~\cite{Shapeev_AL,Shapeev_AL_2018} takes advantage of i) the dimension of features could in principle be significantly lower than the number of degrees of freedom spanned by the molecules (in particular for molecules that are in or close to their equilibrium states) and ii) linearly parametrized local atomistic potential.
Given a set of $K$ molecules, the total energy of the $q$-th molecule could be approximated as $E^{(q)}=\sum_{i=1}^{N} V(\mathbf{x}_i^{(q)}) =\sum_{i=1}^{N} \sum_{j=1}^{m}\theta_j b_j(\mathbf{x}_i^{(q)}) = \sum_{j=1}^{m} \theta_j B_j(\mathbf{x}^{(q)})$
(in matrix form, $\bm{E} = \bm{\theta}\mathbf{B}$),
where $B_j(\mathbf{x}^{(q)}) = \sum_{i=1}^{N} b_j(\mathbf{x}_i^{(q)})$ serves as the effective basis function of dimension $m$ and $b_j(\mathbf{x}_i^{(q)})$ is some function dependent only on the local representation $\mathbf{x}_i^{(q)}$ of the $i$-th atom in $q$,
$N$ is the number of atoms of $q$.
Then deriving the $D$-optimality criteria boils down to finding the ``best'' submatrix (of size $m\times m$) from the over-determined matrix $\mathbf{A}^{K\times m}$ (where $K>m$ and $A_{kl}= B_k(\mathbf{x}^{(l)})$) such that the absolute value of $\rm{det}\mathbf{A}$ reaches its maximum.
Well-established algorithms exist to achieve the $D$-optimality criteria, e.g.~the \texttt{maxvol} algorithm~\cite{maxvol}.
To obtain an optimal set, one typically has to iterate the procedure, one new query per time. If the corresponding magnitude of $\rm{det}\mathbf{A}$ increases, it would be selected (query strategy) and discarded otherwise.
Numerical results~\cite{Shapeev_AL} have shown much improved performance for long-time MD simulation compared to classical on-the-fly learning~\cite{Zhenwei2015}.
However, the downside of $D$-optimality approach is also noticeable, that is, the model has to be updated at each iteration and application of the model could be prohibitive for a data set bearing a large feature space.
Furthermore, the linear potential $B_j$ depends on the proposed representation and the potential form, the latter of which in particular may suffer from lack of expressive power for some systems, i.e., the potential form may lack general applicability for a wide range of molecular systems. And last but not least, this approach relies on the choice of ratio (of $\rm{det}\bm{A}$ values of two consecutive iterations) threshold manually chosen, which has to be tailored for a specific data set and  may not be applicable to other data sets that only differ ever so slightly. 

We note in passing that an alternative view~\cite{Shapeev_AL} of $D$-optimality criteria is to assume that the energy has a Gaussian random noise and the best submatrix $\mathbf{A}$ corresponds to the minimal variance in the solution of $\bm{E} = \bm{\theta}\mathbf{B}$.
Besides, consideration of other properties such as forces could be naturally incorporated into this framework by simply taking derivatives of $B_j$ with respect to Cartesian coordinates, expanding the feature matrix $\mathbf{B}$~\cite{Shapeev_AL}.

Another variance-based approach relies on the GPR directly. That is, once trained, the model can estimate the variance directly, without referring to other criteria (as in the $D$-optimality approach).
The estimated variance serves as a natural indicator telling if any newly added data point would improve the model (if the variance is large with
respect to a user-defined tolerance) or not (if the variance is very small).
A small variance typically also means that the newly added data lie within or close to the current training space, distant otherwise.
Methods like Gaussian process regression (GPR) are stochastic in nature and inherently capable of calculating the variance of prediction. More specifically, GPR aims to estimate the predictive distribution for any test data (unlike the kernel ridge regression model). Related works include that of Snyder et al.'s~\cite{ML4Kieron2012}, in which the Bayesian predictive variance is shown to correlate with the actual error, and recently Reiher's group~\cite{Reiher2019_GPvar_Disp,Reiher2018_GPvar} used GPR to select optimal training sets in an automated fashion to explore chemical reaction network~\cite{Reiher2018_GPvar}, and subsequently adjust for systematic errors in D3-type dispersion corrections, with one (sequential scheme) or multiple systems (batchwise variance-based sampling, BVS) selected each time.

Neural network (NN) based methods also offer a quite distinct perspective on the confidence of predictions.
The general finding is that for NN methods estimates tend to be overconfident~\cite{pmlr-v70-guo17a}, possibly due to the lack of principled uncertainty estimates~\cite{skafte2019reliable} (i.e., NN model typically produces one single value for an input, instead of a preditive distribution like GPR) and/or that the tools for mean estimation perhaps do
not generalize~\cite{skafte2019reliable}.
In spite of the lack of native variance estimate,  variance can still be modeled in practice through consideration of multiple parallel NN models.
In analogy to GPR, uncertainty in NN models can be understood by taking a Bayesian view of the uncertainty of weight with some distribution assumed a priori and then updated by training data.
There exists several variants of such NN models, including the
ensemble neural network models~\cite{peterson_addressing_2017,smith2018less,ciriano_deep_2019,musil_fast_2019} where NN models share the same architecture but varied parameters (typically, ensembles are generated by NN submodels training on distinct subsets of data),
and the dropout
regularized neural network~\cite{gal2016dropout}, a lower cost framework for deriving uncertainty estimates (randomly dropout some nodes each time).
These NN models are highly dependent on the training data, and therefore the predicted variance may not be reliable if the test data is distinct from training data, as is commonly expected for CCS exploration.
Another type of NN model based uncertainty metrics, widely adopted, may alleviate this deficiency, which employs distances in feature space (or some latent space) of
the test data point to the current training data to provide an estimate of similarity measure and
thus model applicability~\cite{Kulik2019_NNvarByDropOut}.
This kind of approach enjoys several other advantages, such as easy
interpretation, model independence as well as potentially fast computation,
but suffer from high dependence on the representation~\cite{Kulik2019_NNvarByDropOut}.
 
\subsection{AMON based QML} 
Having a closer look at all the selection methods presented above, one notices that there is always some footprint of random sampling, 
i.e., one prerequisite for all those methods is a pre-existing starting training data, usually randomly selected, and reaching convergence of training data through iterative addition of new feature inputs may be slow if the starting points barely represent the space spanned by test data.
The AMONs approach~\cite{Amons_nchem2020} attempts to mitigate these shortcomings through selection of the ``optimal'' training set on-the-fly, i.e.~only after having been provided a given  specific query test feature input.
In essence, AMON based QML exploits
the locality of an atom in molecule which allows to reconstruct extensive properties, such as the ground-state energy, 
in some analogy to the nearsightedness of electronic systems~\citep{nearsightedness,stjin2017}.
For the sake of a succinct discussion, we turn our attention to valence saturated system only and we neglect hydrogens.
However, extension to other systems (e.g., system involving radicals, charges, conformational changes, vibrations,
reactions or non-covalent interactions) are also possible~\cite{Amons_nchem2020,OQML}).
Note that throughout the whole process, we are concerned about only heavy atoms.

\begin{figure}
\includegraphics[scale=0.6]{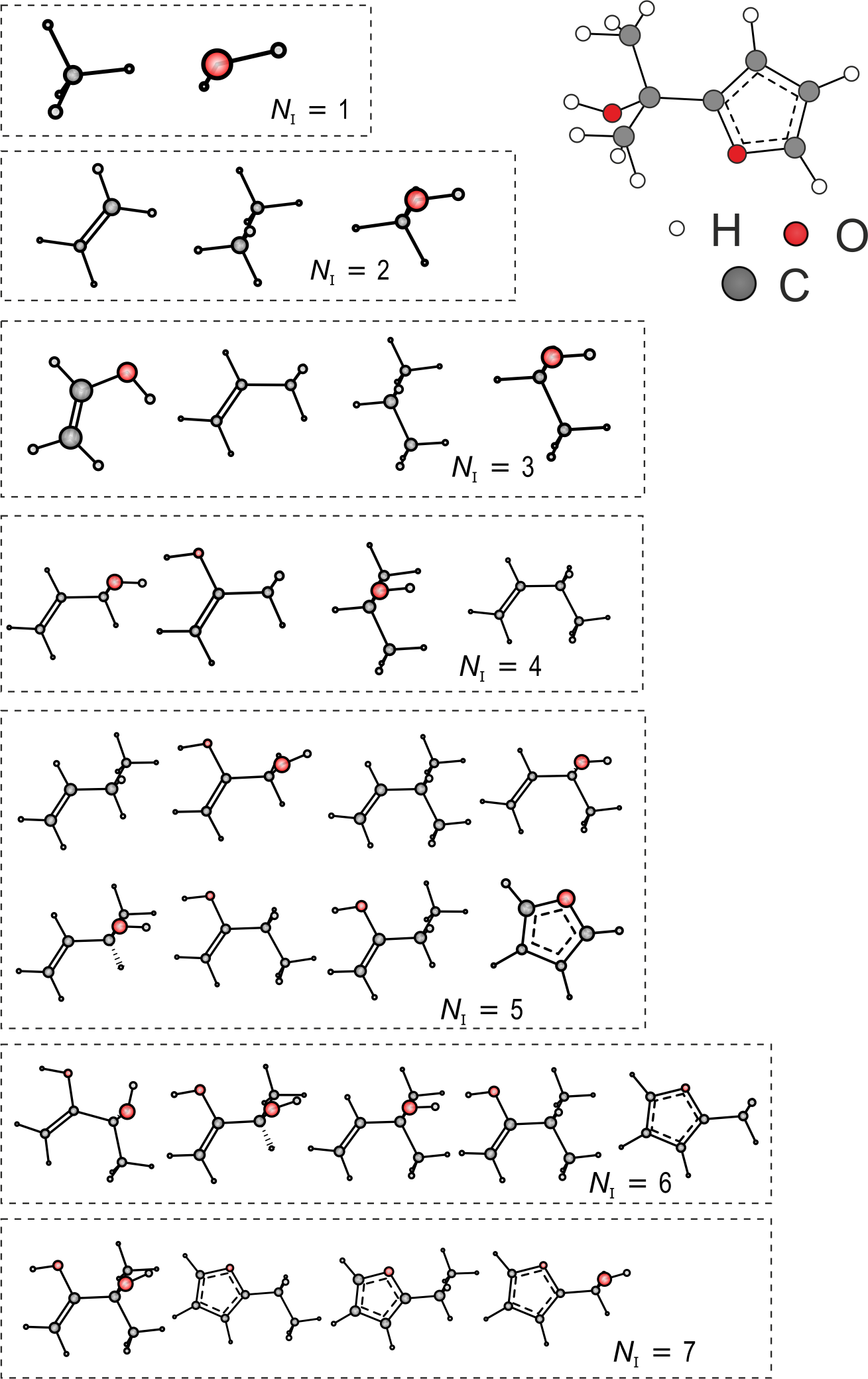}
\caption{\label{fig:amons_example}
	All AMONs size 1 to 7 for training system specific QML models of exemplary query molecule 2-(furan-2-yl)propan-2-ol (top right).}
\end{figure}

The AMONs selection procedure~\cite{Amons_nchem2020} can be divided into four major steps:
i) The connectivity graph $G$ of a query molecule is constructed using its 3D geometry.
ii) Next, all subgraphs are enumerated (the $i$-th subgraph is labeled as $G_i$) of $G$ with increasing number of heavy atoms (denoted as $N_I$). For a given $G_i$, one performs a series of checks to see if it is a representative subgraph: a) is it a connected subgraph? b) does subgraph isomorphism hold true? c) are all atoms valency-saturated after rationalization of the subgraph? And d) is ring structure retained when all associated nodes are present in the current subgraph? If all these criteria are met, then $G_i$ is ready for further filtering and discarded otherwise. Criteria {\bf b} is concise yet informative: subgraph isomorphism ensures that hybridisation states of all atoms in the subgraph are retained, implying that bonds in query graph with bond order larger or equal to 2 are not allowed to break for fragmentation.
iii) Perform geometry relaxation for the corresponding fragment (now with valencies saturated with hydrogen atoms) using, for example, Universal Force Field (UFF)~\citep{uff} or other force-field optimizer with dihedral angles fixed to match the local geometry of the query molecule (to avoid conformational changes in local environments). This step is followed by geometry relaxation using some quantum chemistry program.
At this stage it may happen that the subgraph candidate dissociates (turning into a disconnected graph), or is transformed into a molecule with different connectivity. In the former case, the fragment should be discarded; while in the latter case, the subgraph isomorphism has to be rechecked. 
iv) One proceeds if the subgraph candidate has experienced no change in connectivity, or if subgraph isomorphism is retained despite there is change in connectivity. 
The resulting fragment is selected from the AMON data base. 
 

As the number of atoms in the subgraph increases, one continues looping through \{$G_i$\} until the set has been exhausted. The resulting set of AMONs is considered the query-specific
``optimal'' set which is representative of all local chemistries in the query molecule.

Fig.~\ref{fig:amons_example} shows all AMONs 
for an exemplified QM9 molecule named 2-(furan-2-yl)propan-2-ol
with AMON size ($N_I$) being at most 7 by applying the above algorithm.
\textcolor{red}{Not surprisingly, there exists only two molecules possessing $N_I=1$, i.e., CH$_4$ and H$_2$O.
For $N_I=2$, C=C double bond is allowed to be cleaved from the 5-membered ring, forming a valid AMON H$_2$C=CH$_2$, 
as the resulting AMON retains its original coordination number for C's, meanwhile keeping their valence saturated (i.e., meeting octet rule).
While for fragment like H$_2$C-OH, also extracted from the ring, is not a valid AMON as the valence of C atom is not saturated.
Repeat similar arguments for increasingly larger $N_I$'s, we end up with only 30 AMONs, but which as a whole represent the complete set of local atomic environments present in the target,
and has the potential to extrapolate accurately the properties of the exemplified target QM9 molecule,
as well as infinitely many other molecules that share the same set of AMONs after fragmentation.}

AMON based QML models exhibit improved slopes and off-sets in learning curves, as evinced 
for thousands of molecules after reaching respective training set sizes 
of only $\sim$50 on average. 
By contrast, twenty times larger training set sizes are required 
using random sampling~\cite{Amons_nchem2020}.
One should  note that graph based AMONs are not omnipotent. They are best suited for sampling compositional space of large systems. 
To extend AMONs to also handle configurational spaces is possible in principle, but not trivial as it is also true for extension to systems without explicit graphs, such as metals, metal surfaces, or molecular crystals or liquids.

\begin{figure*}[th!]
\includegraphics[scale=0.8]{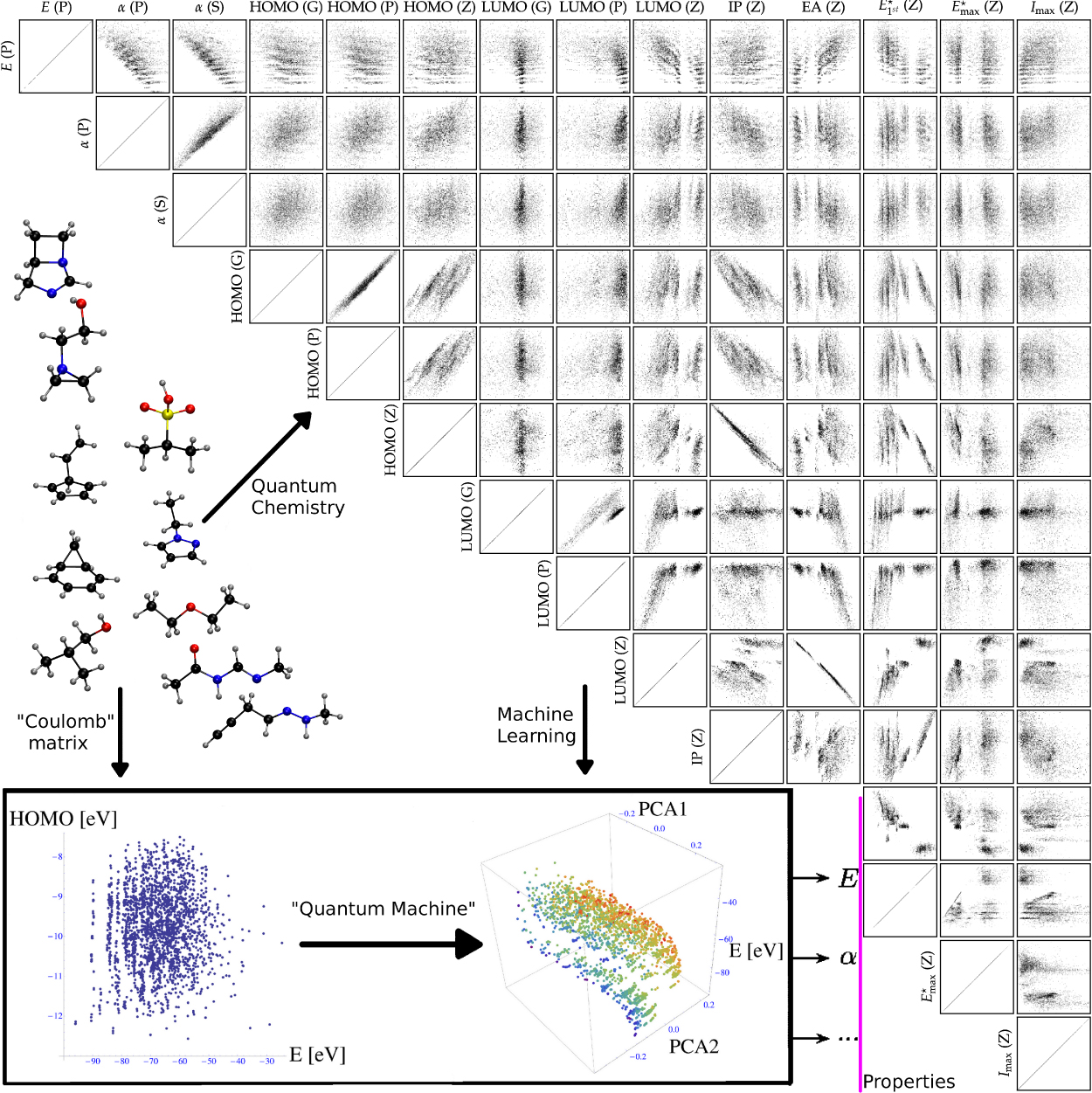}
\caption{\label{fig:NN}
Property vs.~property matrix for $\sim$7k organic molecules at various levels of theory. 
A multi-property neural net trained in CCS encodes underlying correlations as evinced by the first principal components of the last layer for 2k molecules not part of training. 
Reproduced from Montavon, Rupp, Gobre, Vazquez-Mayagoitia, Hansen, Tkatchenko, Müller and von Lilienfeld, New J. Phys., Vol. 15, 095003 (2013)~\cite{Montavon2013};
licensed under a Creative Commons Attribution 3.0 licence.
}
\end{figure*}

\section{Properties}
\textcolor{red}{As we are mainly interested in supervised learning throughout this text, properties (or labels) of molecules have to be always paired with some molecular representation.
Starting from regression of experimental properties, e.g., atomization energies, dipole moment, boiling point, in the early practises of machine learning,
now the scope of properties has been expanded significantly.
}

\textcolor{red}{Due to its determining role for stability and dynamics, energy is among the most important properties, and it is also the primary target property of most studies.}
As early as in 2011~\cite{anatole-MilindDenis2011}, reorganization energies in a sub-space of CCS consisting of polyaromatic hydrocarbons relevant
to photo-voltaic applications were already predicted using ML models. 
While the pioneering work on demonstrating the applicability of QML models 
for navigating CCS was published in 2012 for atomization energies only~\cite{CM}, 
a multi-property neural network was published shortly after~\cite{Montavon2013}, 
covering not only atomization energies but also polarizabilities, molecular orbital eigenvalues, ionization potentials, electron affinities, and excited states properties at various levels of theory.
\textcolor{red}{The correlations among these properties have confirmed some of the well-established physical principles, as well as shown some interesting patterns.}
As illustrated in Fig.~\ref{fig:NN}, \textcolor{red}{
the ionization potential (IP) is well correlated with the HOMO energies, as expected from Koopman’s theorem;
The polarizability is linked to the stability, as insinuated by the hard-soft acid-base principle.
Properties calculated at different levels of theory are strongly correlated, suggesting the possibility to exploit implicit correlations for the training of QML models with superior data-efficiency.}
\textcolor{red}{What is more interesting is that, for properties such as HOMO and atomization energies displaying little correlation,}
after training the neural network encodes some of the underlying and hidden correlations among these properties (box in Fig.~\ref{fig:NN}), indicating already in 2013 that neural network based QML models are amenable to `explainable AI', as also illustrated subsequently in 2017 for effective molecular chemical potentials~\cite{DTNN2017}.

While QML commonly deals with properties which correspond to observables, other well defined but more arbitrary labels can also be modelled.
Examples include atomic charges or energies which do not have a unique definition. 
A more exotic application consists of successfully trained QML models of `time-to-solution' in terms of estimates of the number of iterations necessary to reach convergence for given initial conditions:
In 2020, QML models of the computational cost of common 
quantum chemistry calculations have been demonstrated to enable optimal load-balancing and scheduling in ensemble calculations of high-throughput compute campaigns through CCS~\cite{heinen2020machine}.

\textcolor{red}{To provide a more comprehensive perspective on the interesting subject of property,
below we divide all properties into three main categories depending on the number of atoms/species involved: atomic property (atom/bond/functional group in a molecule), molecular property (the entire molecule), and inter-molecular property (at least two molecular species).
And within each section, we briefly review models of important properties in a rough chronological order.
Note that the boundary between different categories is not clear-cut.
For instance, the highest vibrational frequency of a molecule may be attributed to certain functional group, 
but only in an approximate way,
the exact value of which may still depend on all the other atoms in the molecule.
For this and similar cases, we prefer to classify the relevant properties into atomic, rather than molecular properties.
}


\subsection{Atomic}
\textcolor{red}{Generally speaking, atomic properties are relatively easy to learn
as they typically benefit the most from the general assumption of locality of an atom in a molecule.}
Based on the QM9 data-base, QML models were introduced for atomic properties, such as core level excitations, forces (see previous section), or NMR-shielding constants~\cite{MLatoms_2015}. 
Atomic QML models of electrostatic properties, such as atomic charges, dipole-moments, or atomic polarizabilities were introduced in 2015~\cite{MTP_Tristan2015}, 
and their use for the construction of universal
non-covalent potentials was established in 2018~\cite{bereau2018non}. 
Deep neural networks for similar properties were also contributed in 
2018 and 2019 by Unke and Meuwly~\cite{unke2018reactive,PhysNet2019}.
Information from topological atoms has also been used to build dynamic 
electron correlation QML models in 2017~\cite{mcdonagh2017machine}.
In 2017 and 2018,
atomic energies and potentials were also discussed in
Refs.~\cite{SchuettTkatchenkoMueller2017,Amons,FCHL,chen2018atomic}
QML models of polarizabilities based on tensorial learning were presented 
in 2020~\cite{WilkinsGrisafiRobMichele2020polarizabilityPNAS}, 
and most recently, Gastegger and co-workers introduced external field effects 
within neural networks and demonstrated interesting performance for predictions 
of IR, Raman, and NMR spectra, as well as for continuum solvent effects on 
chemical reactions~\cite{gastegger2020FieldSchNet}. 
Multi-scale models of atomic properties have also been proposed~\cite{grisafi2020multi}.

QML models of NMR shifts in molecules were first studied in 
2015~\cite{MLatoms_2015} and 2017~\cite{Amons_nchem2020}, followed
by shifts in solids in 2018~\cite{paruzzo2018chemical,ceriotti2019bayesianNMR},
NMR shifts in solvated proteins, coupling, a kaggle challenge, 
and an in-depth revisiting in molecules were all contributed in 
2020~\cite{li2020accurate,navarro2020dft,bratholm2020community,gupta2020revvingNMR}.

In 2017, self-correcting KRR based models of potential energy surfaces 
and vibrational states were presented in Ref.~\cite{dral2017structure},
as well as neural network based molecular dynamics for the calculation
of infra-red spectra~\cite{gastegger2017machine}. 
Out of all QML models for properties studied in the 2017 
overview study on the CCS of QM9~\cite{HDAD},
it was only for the highest vibrational (fundamental) frequency that random forests
performed better than KRR or neural networks, 
the likely reason being that the QML model's task consisted `only' of 
detecting if an O-H or N-H bond present on top of the C-H bonds, 
and to assign the typical corresponding bond-frequency,
and that typically random forests work well for such classification tasks.
Other 2018 studies dealing with infrared spectra include
Refs.~\cite{OQML,Isayev2018charges,Nebgen2018charges}.

\subsection{\textcolor{blue}{Molecular}}
\textcolor{red}{At the molecular level, properties are greatly diversified, ranging from properties for ground state to excited ones, from static to dynamic ones,
as well as from single molecule in vacuum to condensed phase, etc.}

\textcolor{red}{Beside energy,} electronic properties for, such as excited states, quantum transport or
correlation, QML models that are applicable throughout CCS, have remained 
rather sparse over the years.
Examples include QML models for electron transmission coefficients for transport across
molecular bridges of varying composition~\cite{QuantumTransportML2014},
and Anderson impurity models~\cite{LF_DMFT_ML2014} in 2014. 
And dynamical mean field theory~\cite{LF_DMFT_ML2015}
and excitation energies~\cite{ML-TDDFTEnrico2015} in 2015.
Only recently QML has been extended to also study nonadiabatic excited states dynamics
for given systems (conformational sampling)
by Dral, Barbatti and Thiel~\cite{dral2018nonadiabatic},
or Westermayr and Marquetand~\cite{westermayr2019machine,westermayr2020neural}.
And the recent introduction of SchNarc~\cite{SchNarc}, 
a combination of the deep neural net architecture SchNet~\cite{SchNet} 
and the surface hopping {\em ab initio} molecular dynamics code 
SHARC~\cite{SHARC}, 
has led to promising first results for CCS studies involving small sets of 
small molecules~\cite{westermayr2020deep}.
For more details and references to this rapidly moving and growing field,
we refer to the recently published reviews 
on this field~\cite{westermayr2020CR,westermayr2020book,westermayr2020mlst}.

QML models of electron affinities and ionization potentials with deep neural networks
have also recently been proposed~\cite{zubatyuk2020teaching}.
Symmetry conserving neural networks for efficient calculations of electronic and
vibrational spectra have also been presented in 2020~\cite{zhang2020efficient}.

\subsection{\textcolor{blue}{Inter-molecular}}
\textcolor{red}{As the system becomes more complicated, the associated properties also tend to show more interesting, and sometimes surprising patterns.
Hereafter, we will focus on energetic properties, unless otherwise stated.
Depending on whether or not the system has experienced significant reconstruction in the relative orientation between atoms,
inter-molecular energetics could be further divided into inter-molecular binding energy or reaction energy/barriers.
Below we summarise relevant contributions for each of these two sub-categories.}

In terms of binding energies within assemblies of atoms, ever since Ref.~\cite{CM},
a large variety of systems has been addressed by now,
reaching from formation energy predictions of diverse inorganic materials
~\cite{Elpasolite_2016,WolvertonMLcrystals2017,FCHL},
over models of chemical bonds in molecules\cite{Parkhill2017bonds},
to 
models of electronic properties of transition metal complexes~\cite{Kulik2017transitionmetals}.
GPR/KRR based QML models have represent a unified approach, 
as demonstrated for applications to surface reconstructions, organic molecules, as well as protein ligands~\cite{CeriottiScienceUnified2017}.
Symmetry adapted learning of tensorial properties was introduced 
in 2018~\cite{grisafi2018symmetry}, as well as
neural networks for atomic energies~\cite{chen2018atomic},
on-the-fly learning for structural relaxation~\cite{jacobsen2018fly},
crystal graph convolution networks for materials properties~\cite{GrossmannCNN2018},
solvation and acidity in complex mixtures~\cite{ClemenceCeriotti2020SolvationAcidity},
and a machine learning based understanding of the chemical
diversity in metal-organic-frameworks~\cite{BerendSmit2020DiversityMOFML}.
A recent review of big data in metal-organic-frameworks was also
published in 2020~\cite{BerendSmit2020BigDataMOFreview}.

\textcolor{red}{Accurate QML prediction of reaction related properties, reaction barrier in particular, poses as a more difficult task, as typically off-equilibrium configurations are involved,
and the training space is undersampled.}

The use of QML models to investigate properties relevant for catalysis
represents another major domain of applications. 
A GPR model was used in 2016 to estimate 
free energies of possible adsorbate coverages for surfaces in order to 
accelerate the construction of Pourbaix diagrams~\cite{NorskovML2016}
In 2017, Ulissi et al. introduced a neural network based exhaustive 
search enabling the identification of active site motifs 
for CO$_2$ reduction~\cite{ulissi2017machine},
as well as a GPR based estimator of adsorption energies 
for identifying the most import reaction step~\cite{ulissi2017address}.
QML models of reaction barriers of elementary reactions 
(using 236 dehydrogenation, 38 N$_2$ dissociation, 41 O$_2$ dissociation examples)
on surfaces were proposed by Singh et al.~in 2019~\cite{singh2019predicting}
Quantum machine learning based design of homogeneous catalyst 
candidates was presented in 2018~\cite{clemence2018catalystML}.
In 2020, QML models of competing reaction barriers and 
transition state geometries corresponding to S$_N$2 and E2 reactions 
in the gas-phase were successfully trained and applied 
throughout a CCS covering thousands of reactants~\cite{heinen2020quantum},
relying on the QMrxn data-set~\cite{QMrxn}.
That same year, Bligaard and co-workers employed active learning to 
identify stable iridium-oxide
polymorphs and study their usefulness for the acidic oxygen 
evolution reaction~\cite{Bligaard2020activelearning},
introduced a Bayesian framework for adsorption energies of bimetallic alloy catalyst candidates~\cite{Bligaard2020bayesian},
and proposed a bond-information based GPR 
as a means to speed up structural relaxation across different types of
atomic systems~\cite{del2020machine}.
In 2020, neural networks have been proposed for the prediction of overpotentials
relevant for heterogeneous catalyst candidates~\cite{keith2020catalystNN},
as well as a higher order correction scheme in alchemical perturbation density functional
theory applications to catalytic activity~\cite{griego2020machinealchemy}.
An overview on machine learning for computational heterogeneous catalysis
was also contributed in 2019~\cite{lamoureux2019machine}.

\section{Data sets}
As implied already in previous sections, 
the availability of training sets is vital for any machine learning. 
Admittedly, it would be ideal to generate training set only when necessary, i.e.~to minimize the number of QM computations throughout CCS, or for converging the sampling using molecular dynamics.
However, for general applications of QML, a pre-existing data set is indispensable,
for instance, to tackle the inverse design problem to identify some compound with unknown composition and exhibiting specified and desirable ground-state physiochemical properties.
Currently, this is only feasible with a given labeled data set being as representative as possible for the local chemistries that we know to affect the properties of interest.

Alongside the increasing popularity of QML in chemistry and related sciences, many data sets have emerged in recent years. By now, there are a multitude, built for various purposes. 
Here we highlight those data sets that encode quantum information throughout CCS, with an overview given in Table~\ref{tab:db}.

\subsection{GDB} \label{sec:gdb}
The synthetic GDB (generated data base) data sets created by Reymond and co-workers for the main purpose of exploring the CCS of organic drug-like molecules comprise the probably largest list of systematically generated molecular graphs (constitutional and compositional isomers only) of small to medium sized molecules~\cite{gdb11,gdb13, gdb17, gdb17_exploration}. 
To date, GDB17~\cite{gdb17, gdb17_exploration} represents the single largest set of molecules, which contains more than 166 billion molecules made up of H, C, N, O, S, and halogens (up to 17 non-hydrogen atoms) obeying certain chemical rules for stability and synthesizability. 
GDB17 has two main subsets, GDB11 (26M)~\cite{gdb11,gdb11_exploration} and GDB13 (970M)~\cite{gdb13}, together with a variety of smaller subsets featuring specificity of organic chemistry. 
Due to its systematic enumeration, interesting new structures have been identified and subsequently been synthesised, as exemplified by the synthesis of trinorbornane~\cite{bizzini2017synthesis}.

Other than the implicit information that any compound listed corresponds to a stable constitutional isomer, the original GDB data sets are unlabeled in the sense that only molecular composition and connectivity information are detailed, without calculated quantum properties.
The first extension of the GDB data set to also include quantum data, QM7~\cite{CM} consists of 7,165 ground state geometries and energies of molecules of up to 23 atoms (with up to 7 heavy atoms C, N, O, or S) calculated at the PBE0 level.
QM7 is also the first quantum benchmark dataset covering the organic subspace CCS for QML. Some extensions exists, such as QM7b~\cite{qm7b}, QM7b multi-level data set~\cite{zaspel2018boosting} (QM7bMl for short) and QM7-X~\cite{qm7x}.

QM7b~\cite{qm7b} extends QM7 by including chlorine-containing molecules (expanding the set size to 7211), and reporting 13 additional calculated electronic properties (e.g.~polarizability, HOMO/LUMO energies, excitation energies). 
QM7bMl~\cite{zaspel2018boosting} was designed for studying QML combinations with legacy quantum chemistry methods such as multi-level, multi-fidelity or transfer learning. 
Starting from the original coordinates at PBE level, geometries of QM7b molecules were refined at the level of B3LYP/6-31G(D) and subsequently single-point energies were calculated at 9 levels of theory, corresponding to all possible combinations of electron correlation treatment \{HF, MP2, CCSD(T)\} and basis sets \{STO-3G, 6-31G, cc-pVDZ\}).
QM7-X, the largest extension of QM7, is a comprehensive dataset comprising $\sim$4.2 M equilibrium and non-equilibrium structures of QM7b molecules, accompanied by 42 physicochemical properties computed at the PBE0+MBD level, covering global (molecular) and local (atom-in-a-molecule) properties ranging from ground state quantities (such as atomization energies and dipole moments) to response quantities (such as polarizability tensors and dispersion coefficients). 

Due to the limited molecular size, QM7 and its variants, are scarcely scattered across CCS, and barely begin to represent its full diversity and complexity. 
Targeting `big data' Ramakrishnan et al.~released the QM9~\cite{QM9} data set in 2014, derived from molecular graphs drawn from GDB17~\cite{ReymondChemicalUniverse3}, totalling $\sim$134k organic molecules made up of C, H, O, N, or F, and up to 9 non-hydrogen atoms.
Except for equilibrium geometries and electronic ground state properties, QM9 also records a series of thermochemical properties at 298 K and 1 atm pressure estimated based on harmonic frequencies, namely enthalpies, and free energies of atomization at the level of
B3LYP/6–31G(2df,p).
Alongside, additional QM data is reported for the subset of all of QM9's 6k constitutional isomers with sum formula  C$_7$H$_{10}$O$_2$, i.e.~thermochemical properties computed at the G4MP2 level.
\textcolor{red}{In 2020, QM9 was augmented by more accurate energies, calculated at multiple levels of theory, including M06-2X, wb97xd and G4MP2.~\cite{narayanan2019_qm9_g4mp2}}
Another similar data set, alchemy data set~\cite{alchemyDB} (sized 119,487) expands the volume and diversity of QM$x$ series and is made up of 9 to 14 C, N, O, F, S and Cl atoms, sampled from the GDB MedChem subset of GDB17~\cite{gdb17}.

The only data set that deals with excited state properties across CCS is QM8~\cite{qm8}, totalling $\sim$20k structures sub-sampled from QM9 and 
comprising up to eight heavy atoms C, O, N, or F. Ground-state energies ($S_0$) and the lowest
two vertical electronic singlet-singlet excitation energies ($S_1$ and $S_2$) are included, calculated at two TDDFT levels employing the density functional theory/basis-set combination PBE0/def2-SVP or CAM-B3LYP/def2-TZVP, as well as post-Hartree-Fock level CC2/def2-TZVP. Corresponding oscillator strengths ($f_1$) for each transition from $S_0$ to $S_1$ have also been recorded.

As also evinced for GDB17, when increasing the number of atoms per molecule the data set quickly grows out of control, and it becomes prohibitive to conduct QM calculations for comprehensive subsets of CCS. The Amon based dictionary of building blocks designed to cover GDB~\cite{ReymondChemicalUniverse3} and Zinc~\cite{zinc} and containing no more than 7 heavy atoms  (AGZ7) has been introduced to alleviate this curse of dimensionality~\cite{agz7}. It was obtained 
 by systematically fragmenting all larger molecules (from GDB17 \& Zinc~\cite{zinc}) into smaller entities containing no more than 7 non-hydrogen atoms (i.e., atom-in-molecule based fragements, aka, AMONs~\cite{Amons_nchem2020}).
To date, AGZ7 is the most compact yet most diverse data set relevant for organic/bio-chemistry, totalling only 140k molecules but covering up to 13 elements
(H, B, C, N, O, F, Si, P, S, Cl, Br, Sn and I). It also includes a similar set of properties as in QM9, but relying on a slightly different level of theory (B3LYP/cc-pVTZ as well as pseudopotentials for Sn and I).

Apart from QM7-X~\cite{qm7x}, all data sets mentioned so far deal with exclusively equilibrium geometries only, representing the typical constraint for what defines a stable molecule. 
In order to enable the QML based study of dynamics and reactivity  of non-equilibrium geometries throughout CCS, however,
configurational sampling involving non-stationary geometries has to be also accounted for through the data sets. 
Similar to QM7-X, ANI-1~\cite{ani1_db} also explores  non-equilibrium geometries, but for relatively larger systems drawn from GDB11~\cite{gdb11,gdb11_exploration}. 
It consists of more than 20M off-equilibrium structures (sampling both chemical and conformational degrees of freedom) and wB97x/6–31G(d) energies for 57,462 small organic molecules containing up to 11 CONF atoms.
Two follow-up data sets expand ANI-1 considerably, i.e., ANI-1x and ANI-1ccx~\cite{ani2_db}.
The former contains multiple QM properties (density-derived properties and forces) from 5M DFT calculations (wB97x/6-31G* \& wB97x/def2-TZVPP), while the latter contains 500k CCSD(T) energies for estimated CBS limits.

For MD simulations, two main data sets are being frequently benchmarked. One is ISO17~\cite{SchNet,SchuettTkatchenkoMueller2017}, containing MD trajectories of 129 molecules randomly drawn from the aforementioned 6k C$_7$O$_2$H$_{10}$ isomers, each comprising 5,000 conformational geometries with total energies and atomic forces calculated at PBE level plus van der Waals correction~\cite{TS-vdW}.
The other is MD-17~\cite{chmiela2018towards,sGDML_db}, which records
energies and forces from ab initio molecular dynamics trajectories (133k to 993k frames) at the DFT/PBE+vdW-TS level of theory at 500 K for eight organic molecules: benzene, uracil, naphthalene, aspirin, salicylic acid, malonaldehyde, ethanol, toluene.
More accurate CCSD(T) energies and forces are also available, but only for ethanol (with basis cc-pVTZ), toluene and malonaldehyde (cc-pVDZ), and CCSD/cc-pVDZ for aspirin. 
Recently, a revised MD-17 data set was published~\cite{christensen2020gradientsloss}, with a lower noise floor in DFT forces thanks to tighter SCF convergence criteria and denser integration grids.
In 2020, G4MP2 benchmarks of organic molecules with up to 14 non-hydrogen atoms were contributed by Dandu et al.~\cite{CurtissLogan2020FCHLQM14}, and resulting QML models were compared and discussed.

\subsection{PubChem amd ZINC} 
While the GDB family currently dominates QML campaigns, GDB compounds resulted from virtual exhaustive graph enumeration campaigns, and contains mostly molecules for which neither thermodynamics stability nor synthesizability has been established.
Within practical applications, such aspects matter for the experimental design and fabrication of new chemical compounds. 
With respect to QML, some theoretically possible local chemical environments may not be viable within the entire molecular framework, and ruling out such possibilities when training could help to further improve data-efficiency and transferability.
PubChem~\cite{pubchem} is an ever-growing open chemistry database hosted at the National Institutes of Health (NIH). As of October 2020, there were over 111 million unique chemical structures records listed together with many a experimental property, as contributed by hundreds of data sources.
To harvest the richness and popularity of this database, Maho Nakata and co-workers lauched the so-called PubChemQC project~\cite{pubchemQC}, consisting of ground state geometries and properties (at B3LYP/6-31G* level), as well as low-lying excited states of approximately four million molecules via time-dependent DFT at the level of B3LYP/6-31+G*.
\textcolor{red}{A PubChemQC derived subset, called pc9,~\cite{glavatskikh2019_pc9} covering over 99k molecules made up of CHONF was published afterwards, and encoded the same set of properties as  QM9.} The full potential of PubChemQC remains yet to be generally explored.

ZINC~\cite{zinc}, yet another competing large database, focuses more on biochemistry, in particular drug design. Quantum calculations on this database per se have not taken place, except for its associated fragment set. That is,
AZ7, a subset of AGZ7~\cite{agz7}, contains all ZINC AMONs of up to 7 non-hydrogen atoms (with optimized geometries and electronic properties, as for AGZ7 described above), could be considered as an effective set covering all local chemistries of ZINC and may serve as a scaffold for building larger drugs through a theoretical approach.

Beside PubChem \& ZINC, there are several other public big databases being exploited within QML.
One of them is the Cambridge Structural Database (CSD)~\cite{CSD}, based on which Stuke et al.~\cite{OE62} reported a diverse benchmark spectroscopy dataset of 61,489 molecules, denoted OE62. Using geometries optimized by PBE plus vdW correction, OE62 provides total energies and orbital eigenvalues at PBE \& PBE0 levels for all molecules in vacuum and at the PBE0 level for a subset of 30,876 molecules in (implicit) water. 
Also based on CSD, Schober et al. ~\cite{schober2016_CSD_charge_mobility} extracted 95,445 molecular crystals thereof and carried out computations on electronic couplings (at the level BLYP and fragment molecular orbital-based DFT) and intramolecular reorganization energies (by QM/MM with an ONIOM-scheme) as two main descriptors for charge mobility, hoping to facilitate the theoretical design and discovery of high mobility organic semiconductors.

\subsection{Barriers and spin}

Quantum data sets on chemical reaction profiles are rather sparse. The QMrxn~\cite{QMrxn} reports calculated quantum properties for S$_N$2 and E2 reactions amounting to 4,466 transition state and 143,200 reactant complex geometries and energies at MP2/6-311G(d) and single point DF-LCCSD/cc-pVTZ
level of theory, respectively.
QMrxn covers the subset of CCS that is spanned by the substituents -NO$_2$, -CN, -CH$_3$, -NH$_2$,
-F, -Cl and -Br as nucleophiles and leaving groups. 
A different dataset featuring elementary reactions comes from Grambow et al.~\cite{grambow_2020_elementary_reaction_db}, totalling 12k organic reactions that involve H, C, N, and O atoms, calculated at the wB97X-D3/def2-TZVP level, with optimized geometries and thermochemical properties for reactants, products, and transition states.

Going beyond mostly singlet state chemistry, Schwilk et al. introduced QMspin~\cite{QMspin}, consisting of $\sim$5k ($\sim$8k) singlet (triplet) state carbenes derived from 4k randomly selected QM9 molecules.
QMspin also contains optimized geometries (B3LYP/def2-TZVP for triplet state and CASSCF(2e,2o)/cc-pVDZ-F12 for singlet state), as well as the singlet-triplet vertical spin gap computed at MRCISD+Q-F12/cc-pVDZ-F12 level of theory. 

For the QML models of the computational cost of typical quantum chemistry computations (measured by the CPU wall time), Heinen et al. reported the QMt dataset~\cite{QMt}, 
consisting of timings of various tasks (single point energy, geometry optimization and transition state search) for thousands of QM9 molecules,
at several levels of theory including B3LYP/def2-TZVP, MP2/6-311G(d), LCCSD(T)/VTZ-F12, CASSCF/VDZ-F12, and MRCISD+Q-F12/VDZ-F12.

\textcolor{red}{Treating non-covalent interaction (NCI) within QML is an interesting and important research subject, with relevant large datasets emerging only as of recently.
Most notably, several collections of NCI datasets have by now become publicly available~\cite{donchev2021_dimer_edisp}, covering 3,700 distinct types of interacting molecule pairs:
i) DES370K, contains interaction energies for more than 370k dimer geometries with NCI energy calculated at the level of CCSD(T)/CBS (MP2(aVTZ, aVQZ) correlation energy is used for extrapolation); 
ii) DES5M, comprises NCI energies calculated using SNS-MP2, for nearly 5M dimer geometries.
The monomers involved include typical organic species, made up of common p-block elements as well as alkali metal ions, most of which containing no more than seven heavy atoms.
}

Data sets including artificial molecules which violate basic principles of chemical bonding may also of great interest for QML, i.e., they may serve the use of ``soft'' labels (see ~\ref{sec:ALML}) where relatively few compounds might more effectively represent CCS than selected many. 
 MB08-165~\cite{ArtificialMolSet4DFT}, proposed by Grimme, exemplifies that idea, relying on systematic constraints rather than uncontrolled chemical biases. Originally, this data set was designed for benchmarking DFT methods. The potential of such ``unbiased'' artificial molecules as soft labels (training set) in QML has yet to be unraveled.

\subsection{Transition metals} 
Transition metal complex (d-block atom/ion center plus ligands, TMC for short) are pervasive and have been widely used and studied. Due to their complicated electronic structure and the resulting higher computational cost (in comparison to typical organic molecules), the effective exploration of the chemical space spanned by TMCs remains a challenge and current efforts into this subspace are constrained to relatively low level of theory, primarily DFTB or DFT method with some small basis. Examples include tmQM~\cite{tmQM} and the TMC data sets~\cite{Kulik2017transitionmetals,Kulik_TMC_NN_2020,Kulik_MultiobjectiveDesing_TMC} from Kulik and co-workers, as described below.

tmQM~\cite{tmQM} contains  geometries and common electronic properties (as for QM9) of 86,665 mononuclear complexes extracted from the Cambridge Structural Database (CSD). tmQM includes Werner, bioinorganic and organometallic complexes based on a large variety of organic ligands and 30 transition metals.
Based on the DFTB(GFN2-xTB) geometry, common quantum electronic properties (orbital energies, dipole meoment and atomic charges) were computed at the TPSSh-D3BJ/def2-SVP level.

The largest and most comprehensive TMC data sets are from Kulic's group, and have been contributed across multiple publications~\cite{Kulik2017transitionmetals,Kulik_TMC_NN_2020,Kulik_MultiobjectiveDesing_TMC}.
Overall, they correspond to combinations of several metal centers (Cr, Mn, Fe, or Co, Ni) and
a wide range of ligands, ranging from weak-field chloride (Cl$^{-}$) to strong-field carbonyl (CO) along with representative intermediate-field ligands and connecting atoms, including S (SCN$^{-}$), N (e.g., NH$_3$), and O (e.g., acetylacetonate).
Calculated properties are primarily energetic, including total energy, high‐ and low‐spin state energy difference ($\Delta E_{H–L}$),
redox potential and solubility in candidate M(II)/M(III) redox couples,
at the level of theory B3LYP/LANL2DZ (6-31G* for ligands) with or without polarizable continuum model (PCM) for solvents.
The total size could reach up to several millions.

\textcolor{red}{Recently introduced metal-organic frameworks (MOF) dataset by Rosen and coworkers,~\cite{rosen2021_mof14k}, called Quantum MOF (QMOF), represent another broad category of metal complexes.
QMOF consists of computed properties (energy, band gap, charge density and density of states) at the PBE-D3(BJ) level of theory, for more than 14,000 experimentally synthesized MOFs,
which are made up chemical elements that span nearly the entire periodic table.}

\subsection{Solid and solid surface} \label{sec:solidDB}
Compared to TMCs, solid and solid surface present a challenge on their own due to the diversity in composition and spatial arrangements, as well as the resulting complexity of electronic structure.
Typically DFT based methods are used for generating large-scale (or high-throughput) data sets for these systems. The most frequently used method is GGA (PBE) or GGA+$U$ with PAW (projected augmented wave) potentials. 
Based on relaxed geometry, associated calculated properties fall into either electronic properties, e.g., cohesive energy, band structure (and derived properties including density of states and band gap) or response properties such as elastic tensor, bulk modulus, thermodynamic properties (vibrational spectra, free energy, specific heat and entropy) within harmonic approximations.

\begin{table*}\label{tab:db}
\caption{Overview: Synthetic quantum data sets in 3 data families of chemical compound space: GDB (generated data base~\cite{ReymondChemicalUniverse,gdb11,ReymondChemicalUniverse3}), TMC (transition metal complexes), and periodic systems (crystalline solids or surfaces). Properties covered include $E$ (total energy (or atomization energy)), $f$ (atomic forces), $q_A$ (atomic charges), $\mu$ (dipole moments), $\alpha$ (polarizability), $\varepsilon$ (eigenvalues), $E^*$ (excitation energy), $f_i$: oscillation strength for transition from ground state to the $i$-th excited state ($i=$1 or 2), $\Delta E_{H-L}$ (high- and low-spin energy difference), $C_6$ ( London dispersion coefficients), 
$P_{\rm{thermo}}$ (thermochemical properties such as internal energies, enthalpy, free energy, and heat capacity); $E_{\rm{ads}}$ (chemisorption energy). Note that `-' in the third column means elements across the periodic table.}
\begin{tabular}{llllllllll} \hline
Family  &  Data set  & Composition  & Size  & Method                & Properties                                    & Year & Notes                            \\ \hline \hline
GDB     &  QM7~\cite{CM_QM7}       & C,O,N,S    & 7165  & PBE0                  & $E$                                           &     2012 &    &                         \\ \cline{2-8}
        &  QM7b~\cite{qm7b}      & C,O,N,S,Cl & 7211  &  PBE0, ZINDO, GW                     &  $E$, $\varepsilon$, $\alpha$, $E^*$, etc.                                             &       2013 &        &                        \\ \cline{2-8}
        &  QM9~\cite{QM9}       & C,O,N,F    & 134k  & B3LYP/6-31G(2df,p)    & \makecell[l]{$E$, $\mu$, $\alpha$, $\varepsilon$,\\$P_{\rm{thermo}}$, etc.}        & 2014       &                             \\  \cline{2-8}
        &  QM8~\cite{qm8}       & C,H,O,N,F    & 20k   & \makecell[l]{TDDFT, CC2/\\def2-TZVP}  &  $E^*$, $f_1, f_2$                         & 2015                    & excited state                    \\  \cline{2-8}
        &  ANI-1~\cite{ani1_db}     & C,O,N,F    & 20M   & w97x/6-31G(D)         & $E$                   &2017                        & off-equilibrium                  \\ \cline{2-8}
        &  QM7bMl~\cite{zaspel2018boosting}    & C,O,N,S,Cl & 7211  & \makecell[l]{\{HF,MP2,CCSD(T)\}/\\ \{sto-3g, 6-31g, cc-pVDZ\}} & $E$            & 2018    & \makecell[l]{Multi-fidelity\\QML}     \\ \cline{2-8}
        & Alchemy~\cite{alchemyDB} & C,N,O,F,S,Cl & 119k & B3LYP/6-31G(2df,p) & \makecell[l]{$E$, $\mu$, $\alpha$, $\varepsilon$,\\$P_{\rm{thermo}}$, etc.}        & 2019       &                             \\  \cline{2-8}
        &  QM7-X~\cite{qm7x}     & C,H,O,N,S,Cl & 4.2M  & PBE0+MBD              & \makecell[l]{$E$, $f$, $\varepsilon$, $\mu$, \\$\alpha$, $q_A$, $C_6$, etc.}    & 2020  & off-equilibrium                  \\ \cline{2-8} 
        &  ANI-1x~\cite{ani2_db}     & C,O,N,F    & 5M   & \makecell[l]{w97x/def2-TZVPP\\ \& CCSD(T)/CBS}         & $E$, $f$, $\mu$, $q_A$, etc.        &2020                                   & off-equilibrium                  \\ \cline{2-8}
       &  AGZ7~\cite{agz7}       & \makecell[l]{ B,C,N,O,F,Si,\\P,S,Cl,Br,Sn,I}    & 140k  & B3LYP/cc-pVTZ    & \makecell[l]{$E$, $\mu$, $\alpha$, $\varepsilon$,\\$P_{\rm{thermo}}$, etc.}          &2020    &                                  \\ \hline \hline
\multirow{2}{*}{TMC}      &  tmQM~\cite{tmQM}      &  \makecell[l]{3$d$, 4$d$ and 5$d$ \\transition metals, \\B,Si,N,P,As,O,\\S,Se,halogens}            & 86k & TPSSh-D3BJ/def2-SVP   &  \makecell[l]{$E$, $\mu$, $q_A$, $\varepsilon$, etc.} & 2020 & \makecell[l]{GFN2-xTB\\geometry} \\ \cline{2-8}
        &  \makecell[l]{Kulik (MIT) \\~\cite{Kulik_2017_TMC_space,Kulik_MultiobjectiveDesing_TMC}} & \makecell[l]{Cr,Fe,Mn,Co,Ni,\\C,N,O,S,Cl} & $>$2M   & \makecell[l]{B3LYP/LANL2DZ\\(6-31g*)}  &  \makecell[l]{$E$, $\Delta E_{H-L}$,\\ redox potential}  &2017,2020 &                                  \\  \hline \hline
periodic                 &  \makecell[l]{Materials\\Project~\cite{MaterialsProject}}  &  -           &  $>$600k &      PBE                                  &   \makecell[l]{$E$, electronic and\\ response properties}           &                 2011- &              \\    \cline{2-8}
&  AFlow~\cite{AFLOW}                             &  -           &   3M  &      PBE                                  &    \makecell[l]{$E$, electronic and\\ response properties}  &    2012- &     \\  \cline{2-8}
               &  OQMD~\cite{OQMD}                              &  -           &  300k &      PBE                                  &     \makecell[l]{$E$, electronic and\\ response properties}       &   2013- &                            \\   \cline{2-8}
               &  OC20~\cite{oc20}                              &  -  &  $>$1M  &      RPBE                                 &      $E$, $E_{ads}$        &                2020 &               \\ \hline
\end{tabular}
\end{table*}


Relevant well-known solid databases and compute platforms include i) AFlow~\cite{AFLOW}, an open data set of more than 3M material compounds (including alloys, intermetallics and inorganic compounds) with over 596M calculated properties.
ii) The Open Quantum Materials database~\cite{OQMD} (OQMD), a high-throughput database currently consisting of nearly 300k total energy calculations of compounds from the Inorganic Crystal Structure Database (ICSD).
iii) The Materials Project~\cite{MaterialsProject} (\texttt{www.materialsproject.org}) covers the properties of almost all known inorganic materials, currently contains over 131k inorganic compounds and more than 530k nonporous materials;
iv) The Materials Cloud (\texttt{www.materialscloud.org
})~\cite{MaterialsCloud}, a platform designed to enable open and seamless sharing of resources for computational science, driven by applications in materials modelling.
v) The Novel Materials Discovery (NoMaD, \texttt{http://nomad-repository.eu}), led by M. Scheffler, C. Draxl, et al. and vi) the Open Materials Database (\texttt{http://openmaterialsdb.se}, currently under development) spearheaded by R. Armiento. The latter both are public archives for hosting, sharing and reusing material data in their raw form.
Apart from comprehensive public repositories for solid data sets, there are also select contributions for select materials classes, including the aforementioned data set of $\sim$10k AB$_2$C$_2$ elpasolites covering all main-group elements up to Bi from Faber et al.~\cite{Elpasolite_2016}

Regarding solid surfaces, the new Open Catalyst Project~\cite{oc20} aims to help discover and design new catalysts for renewable energy storage using ML (\texttt{https://opencatalystproject.org}), currently including mainly the OC20 dataset~\cite{oc20}, consisting of $>$1M relaxations (over 26M single point evaluations) at RPBE level for a wide range of adsorbates (C-, N- and O-containing species) and surfaces.

\section{Software packages}
To perform and supplement the aforementioned studies with methods and data sets, numerous software-packages have been developed over recent years.
We briefly mention the available codes and categorize them into three main types, the first of which being those related to the
acceleration of legacy quantum codes,
such as {\em ab initio} molecular dynamics (MD) runs in VASP~\cite{jinnouchi2020fly},
Gaussian process based geometry optimization in ASE~\cite{GPminASE},
machine learning adaptive basis-sets within CP2K~\cite{schuett2018MLbasisCP2K},
as well as SNAP~\cite{Thompson2015_SNAP} in LAMMPS, a machine-learning interatomic potential using bispectrum components to characterize the local neighborhood of each atom of the system.

Codes which fall into the second category are standalone packages, some of which having also been interfaced to other atomistic simulation software.
QMLcode~\cite{qmlcode,FCHL19} which is an open-source python-based package featuring the Coulomb-matrix~\cite{CM}, BoB~\cite{BoB}, (a)SLATM~\cite{Amons_nchem2020}, FCHL18 and FCHL19~\cite{FCHL,FCHL19} and other representations.
AQML code~\cite{aqmlcode} is a variant of QMLcode featuring the BAML~\cite{BAML} representation, and on-the-fly selection of AMONs for training~\cite{Amons_nchem2020}.
PLUMED~\cite{bonomi2019_PLUMED} is an open-source, community-developed library that provides a wide range of methods including enhanced-sampling algorithms,
free-energy methods, and MD data analysis capabilities. It also interfaces with some of the most popular MD engines.
TensorMol~\cite{yao2018_tensormol} is a package of neural networks for chemistry, capable of running many common tasks in quantum chemistry such geometry optimizations, molecular dynamics, Monte Carlo, nudged elastic band calculations, etc.. It can also take into account screened long-range electrostatic and van der Waals interactions. 
TorchANI~\cite{gao2020_torchani} is a PyTorch implementation of ANI. It can compute molecular energies, gradients, Hessian and derived properties from the 3D coordinates of molecules. It also include tools to work with ANI datasets (e.g. ANI-1, ANI-1x, etc.).

The third category of software packages deals predominantly with data set construction, management and analysis. In particular, specific platforms include
AFlow~\cite{AFLOW} which has been mentioned above in section~\ref{sec:solidDB}, and AiiDA~\cite{huber2020_aiida}, an open-source
infrastructure for automation, management, sharing and reproduction of
the workflows associated with big data in computational sciences.

\section{Compound discovery}
\textcolor{red}{The computational design and discovery of new compounds can be generally conducted at two distinct approaches.
The Edisonian and more basic one is straightforward, within a brute-force  
high-throughput screening,
through solving Schr\"{o}dinger equations sequentially or in parallel 
for potential materials candidates one by one, followed by subsequent
ranking and selection.
Given sufficient coverage and having used the data for training, the ab-initio solver could successively be replaced by QML models,
capable of making faster and equally accurate predictions of target properties of interest.
It is obvious that such an approach suffers from limited domains of compounds conceived in the first place,
no matter what solver is used for computation of properties.
Also, as the intended search domain expands in CCS, 
the number of possible potential candidates will grow combinatorially.
Therefore, when adopting this strategy one
should refrain from generally expanding the search domain, and rather focus on a constrained sub-domain of compounds,
sharing one or more common features,
e.g., the same stoichiometry and space group,
as was exemplified for the elpasolite family ABC$_2$D$_6$ by Faber, et al.~\cite{Elpasolite_2016},
where compounds with exotic chemical property were identified.}

\textcolor{red}{The second more sophisticated approach attempts to solve the problem 
in an inverse fashion. More specifically, given a specific (range of) value(s) for the target property,
how to best locate the corresponding optimal (set of) compound(s) from CCS.
One particularly promising variant is the gradient-based inverse design~\cite{batista2019designMLalchemy},
which can be reformulated as a global optimization problem,
and has the potential to search chemical subspace with significantly spanned domain,
due to its analytical nature.
Strictly speaking, almost all current ML-guided studies (mostly neural network based) on gradient-based inverse design (e.g., Refs~\cite{chemical_design_VAE}, for a review see~\cite{sanchez2018inverse}) fall into the QSPR regime,
as the input is seldomly 3D geometry, but rather SMILES or other molecular graph derived features (therefore,
the mapping from representation to property is not unique).
This strategy is however the only attainable way by now, as otherwise i)
the search subspace (when optimizing for the ``optimal'' compound)
would become overwhelmingly large due to the explosion of conformational degrees of freedom (Levinthal's paradox);
ii) There exists, to the best of our knowledge, no 3D geometry-based representation that is compact enough for decoding,
i.e., restoring the original geometry from
its representation vector/matrix/tensor (or simply $\mathbf{x}$),
even with the help of a neural network model like variational encoder (VAE),
as the entries in $\mathbf{x}$ are highly intertwined
(significantly more so than the SMILES string).
The fact that many representations are still being haunted by the uniqueness issue,
further plagues these efforts,
as often only two and three-body terms are included in distribution-based representation.
While inclusion of four-body terms are mandatory for reconstructing geometry,
as evinced by the Z-matrix representation of geometry,
the resulting $\mathbf{x}$ would become very expensive for generation,
and more importantly, this could further perplex the feature vector decoder.
However, representing a molecule in its most native form \{Z; R\}, 
i.e.~by the variables employed in the electronic Hamiltonian,
or some transformed form, such as an external potential,
one is free from such problems.
This strategy would be consistent with the aforementioned GCE and LCAP approach detailed in section~\ref{sec:FPCCS}.
}

\section{Outlook and conclusion}
\textcolor{red}{While QML is still in its infancy, very encouraging progress has already been achieved. It is still a long way, however, before we will reach the goal of routinely design and discover novel molecules and materials on a computer. 
Some of the most fundamental problems, also among the most common tasks in quantum chemistry calculations, such as correctly predicting ground state energy and forces of novel molecules or materials with high efficiency and accuracy still remains unresolved at large. Such seemingly simple tasks are particularly challenging when it comes to  systems that are highly distorted, charged, or multi-reference in nature, or that involve long-range non-bonded interactions. Successful QML models could easily demonstrate their applicability by energy ranking of competing structures of real materials. We believe that such tasks will be crucial for subsequent more challenging QML applications. }

\textcolor{red}{Another interesting path to prusue might be the integration of alchemical perturbation theory  into QML. Since the alchemical problem could be essentially reformulated as a ML problem that involves  both energy and energy gradient with respect to nuclear charges. A corresponding extension would exploit similarities between alchemical interpolations in pseudo-potential parameter space and compositional representations that explicitly account for group and period in the periodic table, on top of all the structural degrees of freedom. 
Within the FCHL representation~\cite{FCHL}, preliminary results for inferring properties of chemical elements absent in training have already been obtained (see Fig.~\ref{fig:FCHL}). 
}

\textcolor{red}{Besides the curse of dimensionality, imposed by the compositional, constitutional, and conformational diversity of CCS,
the lack of a more theoretical underpinning of
the genesis of data-sets is maybe among the most severe shortcomings. 
Little, if nothing, is known about fundamental questions such as
i) Are there any basic quantities characterizing the completenes of a molecular dataset, for instance in terms of diversity and/or sparsity? 
ii) Based on the inherent properties of the dataset, representation and regressor, can we infer the performance of a model without actual training/test runs, as translated into the slope and offset of resulting learning curve.
iii) What, if any, characterizes the `correct' distribution in CCS.
Answering such questions rigorously, i.e. based on the laws of physical chemistry (in particular simplified quantum chemical models, providing profuse insights), is not only of conceptual importance, but would also benefit the practical design of more efficient/accurate QML models.
}

\textcolor{red}{Other unresolved issues include i) the lack of appropriate QML model to deal with intensive properties such as HOMO/LUMO energy, or dipole-moments, which may require careful consideration of both local and long-ranged features of a molecule.
ii) the lack of high-accuracy datasets at the level of experimental quality (e.g., CCSD(T)-F12/CVQZ-F12 or multi-reference) for medium-sized molecules (published datasets of such quality are still limited to very few or small molecules, containing typically no more than three heavy atoms).
}

As the field has been growing massively and rapidly~\cite{Burke2020retrospective}, 
we can unfortunately not guarantee completeness of our outlook. 
Furthermore, several related important new research directions, i.e.~going beyond the mere
supervised learning problem of the electronic Schr\"odinger equation,
\textcolor{red}{but may fall out of the scope of `conventional' QML,}
have not been mentioned.
They include, for example, variational auto-encoders 
which can be used 
\textcolor{blue}{to help} solving the inverse
design challenge in CCS (e.g.~applied to the design of improved 
molecular electronics~\cite{gomez2018automatic}),
the reconstruction of quantum states~\cite{Juan2019quantumstates}, 
or to molecular structure generation~\cite{Roth2020MolNet}.
Other intriguing efforts deal with
tackling the problem of reaction planning~\cite{segler2018planning,schwaller2018found,nair2019data,schwaller2019molecular,schwaller2020predicting,pesciullesi2020transfer}, 
phase diagrams~\cite{Juan2017phases,Juan2017fermionphases,Juan2017quantumphases,Roman2018phaseTransitions,Bingqing2019water,Bingqing2020hydrogen},
studying the electronic structure in more depth and detail~\cite{QMB_ANN2017,Kieron2020MLDFT,SchNorb,hermann2020deep,Juan2020quantummatter},
or the systematic 
incorporation of experimental information in order to 
improve experimental design~\cite{ShrierNorquist2016ML4failedExperiments}.

To recap, we have provided succinct explanations and pointers to \textcolor{red}{three major ingredients of QML: representation, regressor, and training set. We have briefly discussed select} relevant studies which deal with the development and use of surrogate machine learning models of quantum properties throughout CCS.
\textcolor{red}{One of the primary goals of QML, i.e., rational computational discovery and design of compounds with desired properties,
however, has not yet been achieved in general, and most of the relevant studies are either conducted in a
high-throughput fashion, merely accelerated by QML, or rely on coarsening the problem through neglect of relevant degrees of freedom.
We have pointed out several open questions and challenges that must be overcome to reach this general goal,
as well as potential solutions, and suggestions about interesting new research directions.
Given the overall rapid growth and the multiple success-cases already achieved in this young field, we are optimistic about its future,
and strongly believe that QML will develop into a helpful component for solving some of the long-standing problems in the atomistic sciences.}

\section*{Acknowledgement}
O.A.v.L. acknowledges support from the Swiss National Science foundation (407540\_167186 NFP 75 Big Data) and from the European Research Council (ERC-CoG grant QML and H2020 projects BIG-MAP and TREX). 
This project has received funding from the European Union's Horizon 2020 research and
innovation programme under Grant Agreements \#952165
and \#957189.
This project has received funding from the European Research Council (ERC) under the European Union’s Horizon 2020 research and innovation programme (grant agreement No. 772834).
This result only reflects the
author's view and the EU is not responsible for any use that may be made of the
information it contains.
This work was partly supported by the NCCR MARVEL, funded by the Swiss National Science Foundation. 
Both authors thank J.~Wagner and F.~A.~Faber for helping with the design of Figs.~\ref{fig:Constellation} and \ref{fig:kernel}.

\bibliography{ref1} 
\bibliographystyle{ieeetr}

\section*{Biographies}
Dr. Bing Huang (Hubei, China, 1987) was initially trained in physical
chemistry under the supervision of Prof. Lin Zhuang in
Wuhan University and completed his Ph.D. there in
2015, investigating and developing reactivity theory
concerning solid surface. Afterwards, he moved to Basel, Switzerland to work as a postdoc with Anatole von Lilienfeld at
Department of Chemistry, University of Basel, shifting research interests to the development of machine learning models and methods in quantum chemistry to
explore chemical compound space. 
As of 2020, he has relocated to Vienna, Austria, to continue his postdoctoral research with Anatole von Lilienfeld at the Faculty of Physics, University of Vienna.
His research interests include electronic structure theory, chemical reactivity theory,
theoretical surface science and quantum machine learning.

O. Anatole von Lilienfeld (Rochester, Minnesota, USA, 1976) is a full university professor of computational materials discovery at the Faculty of Physics at the University of Vienna. Research in his laboratory deals with the development of improved methods for a first principles based understanding of chemical compound space using perturbation theory, machine learning, and high-performance computing.\\
Previously, he was an associate and assistant professor at the University of Basel, Switzerland, and at the Free University of Brussels, Belgium. 
From 2007 to 2013, he worked for Argonne and Sandia National Laboratories 
after postdoctoral studies with Mark Tuckerman at New York University and at the Institute for Pure and Applied Mathematics at the University of California Los Angeles. In 2005, he was awarded a PhD in computational chemistry from EPF Lausanne under the guidance of Ursula R\"othlisberger. His diploma thesis work was done at ETH Zürich with Martin Quack and the University of Cambridge with Nicholas Handy. He studied chemistry at ETH Zürich, the Ecole de Chimie Polymers et Materiaux in Strasbourg, and the University of Leipzig.\\
He serves as editor in chief of the IOP journal {\em Machine Learning: Science and Technology} and on the editorial board of {\em Science Advances}. 
He has been on the editorial board of {\em Nature}'s {\em Scientific Data} from 2014 to 2019.
He was the chair of the long IPAM UCLA program ``Navigating Chemical Compound Space for Materials and Bio Design'' which took place in 2011.
\\
He is the recipient of multiple awards including the Swiss National Science foundation postdoctoral grant (2005), Harry S. Truman postdoctoral fellowship (2007), Thomas Kuhn Paradigm Shift award (2013), Swiss National Science professor fellowship (2013), Odysseus grant from Flemish Science foundation (2016), ERC consolidator grant (2017), and Feynman Prize in Nanotechnology (2018).

\end{document}